\documentclass[a4paper,11pt]{article}
\usepackage{jheppub}
\usepackage{amsmath,amssymb,booktabs,graphicx,microtype,xcolor}
\usepackage{placeins}

\let\originaltableofcontents\tableofcontents
\renewcommand{\tableofcontents}{\begingroup\small\originaltableofcontents\endgroup}
\hypersetup{pdftitle={Full-event anomaly detection for new physics searches at the Electron-Ion Collider},pdfauthor={Athanasakos, Grieninger, Liu, Mangan, Ringer, Szafron}}
\graphicspath{{figures/}}
\newcommand{\searchFigureWidth}{.82\linewidth}

\newcommand{\mjj}{M_{12}}
\newcommand{\mhnl}{M_{e j_2 j_3}}
\newcommand{\mres}{M_{\mathrm{res}}}
\newcommand{\epsS}{\epsilon_S}
\newcommand{\epsB}{\epsilon_B}

\newcommand{\SIC}{\mathrm{SIC}}
\newcommand{\PYTHIA}{\textsc{Pythia}}
\newcommand{\GeV}{\mathrm{GeV}}
\preprint{IQuS@UW-21-134, NT@UW-26-22}
\title{Full-event anomaly detection for new physics searches at the Electron-Ion Collider}
\author[a]{Dimitrios Athanasakos,}
\emailAdd{athanasakosdi@gmail.com}
\author[b]{Sebastian Grieninger,}
\emailAdd{segrie@uw.edu}
\author[c]{Hongkai Liu,}
\emailAdd{hliu6@bnl.gov}
\author[a]{Tymothy Mangan,}
\emailAdd{tymothy.mangan@stonybrook.edu}
\author[a]{\\[4pt]\mbox{Felix Ringer},}
\emailAdd{felix.ringer@stonybrook.edu}
\author[c]{Robert Szafron}
\emailAdd{rszafron@bnl.gov}
\affiliation[a]{Department of Physics and Astronomy, Stony Brook University, Stony Brook, NY 11794, USA}
\affiliation[b]{InQubator for Quantum Simulation (IQuS), Department of Physics, University of Washington, Seattle, WA 98195, USA}
\affiliation[c]{High Energy Theory Group, Physics Department, Brookhaven National Laboratory, Upton, NY 11973, USA}

\abstract{
The future Electron-Ion Collider (EIC) will offer new opportunities to search for physics beyond the Standard Model. We study resonant anomaly detection using machine learning and the full particle content of an event. We consider a new neutral gauge boson and a heavy neutral lepton as illustrative signals with different electron-jet topologies. Weakly supervised learning uses this event information to enhance the sensitivity of an invariant-mass search without identifying individual signal events during training. Comparisons with high-level observables show that the electron kinematics account for an important part of the separation, while correlations among the particles provide further sensitivity. We also construct a conditional full-event generator by adapting a model pretrained on LHC jets to produce EIC  background events. A scan over signal fractions demonstrates substantial signal enhancement using the learned background reference, providing a proof of concept for this approach at the EIC. Finally, we consider a fully unsupervised graph autoencoder that provides modest signal enhancement but retains sensitivity at signal fractions too small for effective weak supervision. Our results motivate full-event anomaly detection as part of the EIC physics program and identify directions for developing methods suited to its distinct kinematics.
}
\keywords{Electron-Ion Collider, Beyond Standard Model, Anomaly Detection, Machine Learning}
\begin{document}
\maketitle
\clearpage

\section{Introduction}
\label{sec:introduction}

The Standard Model describes a wide range of particle interactions, yet leaves fundamental questions about the nature of dark matter, the origin of mass, and the matter-antimatter asymmetry unanswered. Physics beyond the Standard Model (BSM) may address these questions through models with new particles and interactions whose masses and couplings are not known in advance. Most collider searches are optimized for a specific signal hypothesis. This usually yields the best sensitivity to that hypothesis, but limits the range of signatures a single analysis can cover. Anomaly detection offers a complementary strategy by identifying departures from the expected background without specifying the detailed signal distribution. For resonant signals, i.e.\ those involving on-shell production of new particles, we can retain an invariant mass as the search variable and use the remaining event information to sharpen a possible excess. Such methods have been developed extensively for the LHC~\cite{Collins:2018epr,Golling:2023yjq,Belis:2023mqs} and applied to data by ATLAS~\cite{ATLAS:2020iwa,ATLAS:2023azi,ATLAS:2025obc} and CMS~\cite{CMS:2024nsz}.

An example is a boosted resonance decaying into two quarks that form a single jet with two hard prongs. Its internal structure differs from that of a typical quantum chromodynamics (QCD) jet initiated by a single quark or gluon, which is dominated by one hard prong. The LHC Olympics used this contrast to create a public benchmark: a heavy resonance decays to two lighter particles, each producing a two-prong jet, compared to a QCD dijet background~\cite{Kasieczka:2021xcg}. The challenge supplied both labeled development samples and samples with hidden signals, allowing a broad range of anomaly-detection methods to be compared. This raises the question of how well methods developed for such distinctive jet substructures transfer to the different final states and kinematics of the future Electron-Ion Collider (EIC).

The EIC will combine polarized electron and ion beams, high luminosity, asymmetric collision kinematics, and broad particle-identification capabilities~\cite{AbdulKhalek:2021gbh,Accardi:2012qut, Alexandrou:2026xrb}. Its primary physics program targets the structure of nucleons and nuclei, but its experimental capabilities additionally offer new opportunities for searches for BSM physics~\cite{Boughezal:2020uwq,Boughezal:2022pmb,AbdulKhalek:2022hcn,Li:2021uww,Yan:2021htf,Jiang:2025frv,Wen:2024cfu}. Electron-proton collisions have a long history as probes of new physics~\cite{South:2016cmx}. H1 and ZEUS at HERA carried out targeted searches over a wide range of models~\cite{H1:2011ije, ZEUS:2012pwm,ZEUS:2016ygq, H1:2004rlm,H1:2008aak}. At the EIC, the scattered lepton constrains the hard interaction in neutral-current deep-inelastic scattering (DIS), and hadronic final states become accessible at relatively low scales. Together, these features make the EIC's new-physics program complementary to searches at the LHC.

Machine learning allows us to use the full particle content of the event rather than a small set of high-level observables or summary statistics. Its role in collider physics spans reconstruction, classification, and inference~\cite{Larkoski:2017jix, Guest:2018yhq, Albertsson:2018maf,Radovic:2018dip,Plehn:2022ftl,Karagiorgi:2021ngt}. Measurements at HERA and studies for the EIC have demonstrated the value of particle-level information for unfolding, DIS reconstruction, and identifying the underlying hard process \cite{Andreassen:2019cjw, H1:2021wkz, Arratia:2021tsq, Lee:2022kdn,Allaire:2023fgp}. The EIC provides a distinctive setting for these methods. Its asymmetric beams, moderate transverse momenta, and a fully measured scattered electron give access to correlations across the hadronic final state~\cite{Aschenauer:2019kzf,Arratia:2019vju,Page:2019gbf}. We therefore compare full-event learning against summary statistics that describe the electron, the overall event activity, and the distribution of particles.

We study a neutral gauge boson associated with a gauged baryon-minus-lepton ($B-L$) symmetry and a heavy neutral lepton (HNL) as representative BSM signals~\cite{Basso:2008iv,Abdullahi:2022jlv}. Both processes yield a final state electron and several resolved jets, but the electron plays a different role in each. In the B-L scenario, the electron is the scattered beam electron that radiates the neutral gauge boson, whereas in HNL production, the incoming electron is absorbed into the HNL through charged-current scattering and re-emerges only when the HNL decays.
The first topology probes the recoil against a radiated resonance, while the second probes the decay kinematics of a heavy particle. These examples connect anomaly detection to the wider EIC program of searches for new neutral bosons~\cite{Liu:2021lan,Yan:2022npz,Davoudiasl:2023pkq,Balkin:2023gya,Gao:2024rgl,Davoudiasl:2025rpn,Adhikary:2026rck,Balkin:2025rtc,Balkin:2026whv,Davoudiasl:2026rqa}, charged-lepton flavor violation \cite{Gonderinger:2010yn, Cirigliano:2021img, Davoudiasl:2021mjy,Zhang:2022zuz,Davoudiasl:2024vje,UrrutiaQuiroga:2026yee}, and tests of lepton flavor universality~\cite{Deng:2025hio}.

These benchmarks allow us to ask to what extent methods developed for the LHC carry over to EIC events. First, can learning from the complete event reveal a resonance that would be difficult to identify from its mass alone? We train a classifier to distinguish event samples with different signal fractions. Classification without labels (CWoLa) makes it possible to recover signal-sensitive information from these mixed samples without requiring a model of what the signal distribution looks like~\cite{Dery:2017fap,Cohen:2017exh,Metodiev:2017vrx,Komiske:2018oaa}. We adapt a jet-pretrained model to EIC events~\cite{Mikuni:2024qsr,Bhimji:2025isp} and compare its sensitivity with that obtained from conventional summary statistics. 

Second, how much of this sensitivity survives when the background must be learned from data? An independent sample drawn from the correct simulated distribution allows us to isolate the information available to the classifier. Estimating that reference from control regions introduces an additional uncertainty. We build on sideband-based background estimation~\cite{Nachman:2020lpy, Andreassen:2020nkr, Hallin:2021wme,Raine:2022hht,Buhmann:2023pmh} and full-event generation at the EIC~\cite{Devlin:2023jzp,Araz:2024bom} to construct a conditional model. A matched scan over signal fractions then quantifies the cost of this substitution by training the same classifier with either an ideal or a learned background reference.

Third, we investigate whether fully unsupervised anomaly-detection techniques can enhance resonance searches at the EIC. We consider an autoencoder, which learns to reconstruct background-dominated events and assigns each event an anomaly score based on its reconstruction error~\cite{Heimel:2018mkt,Farina:2018fyg,Cerri:2018anq,Blance:2019ibf}. We extend the sparse graph-autoencoder approach developed for LHC jets in Ref.~\cite{Araz:2026graph} to full EIC events. It can be trained when the signal component is too small to provide useful weak labels, making it complementary to weak supervision~\cite{Collins:2021nxn}. We assess how effectively the reconstruction score distinguishes signal from background and improves the search sensitivity. Other model-agnostic approaches use learned likelihood ratios, density estimation, factorization, or iterative refinement of weak labels~\cite{Dagnolo:2018cun,Hajer:2018kqm,Metodiev:2023izu,Amram:2020ykb}.

For the $B-L$ benchmark, we find that weakly supervised searches exploit the electron recoil together with correlations in the hadronic final state to substantially improve the sensitivity of a mass-based search. The HNL benchmark probes the complementary case in which the electron belongs to the resonance decay, so its correlations with the decay jets provide additional ways to distinguish signal from DIS background. The learned background preserves substantial signal enrichment across the resolved part of a matched signal-fraction scan, providing a proof of concept for the same search with a reference learned from sidebands. The autoencoder also separates the $B-L$ signal from the background, but its reconstruction score achieves a lower signal enhancement score. A plausible explanation is that signal and background share particle patterns that the autoencoder can reconstruct with similar accuracy, even though those patterns occur at different rates in the two samples. A classifier can exploit these differences in frequency without requiring signal events to have larger reconstruction errors. Our findings suggest several directions for developing new anomaly-detection methods tailored to the physics of the EIC.

The remainder of this paper is organized as follows. Section~\ref{sec:physics} introduces the physics benchmarks and event reconstruction. We describe weak supervision and classifiers in section~\ref{sec:learning}, followed by the conversion of their scores into a resonance significance in section~\ref{sec:inference}. Section~\ref{sec:results} presents the classifier results. Sections~\ref{sec:generator} and~\ref{sec:ae} discuss the conditional generative model and the fully unsupervised autoencoder, respectively. We conclude in section~\ref{sec:conclusions}.

\section{Physics benchmarks and event reconstruction}
\label{sec:physics}

\subsection{BSM and background processes}

We use two illustrative BSM benchmarks with different electron-jet topologies to study how full-event anomaly detection can enhance resonance searches at the EIC. The first is a neutral gauge boson associated with a gauged $B-L$ symmetry. Among all the accidental global symmetries of the Standard Model, the combination of baryon and lepton number $B-L$ is unique because it can be gauged: adding one right-handed neutrino per generation cancels the associated anomalies, whereas $B$ and $L$ separately remain anomalous~\cite{Davidson:1978pm, Marshak:1979fm,Mohapatra:1980qe}. Those same right-handed neutrinos acquire Majorana masses when $B-L$ is broken, so gauging $B-L$ ties the existence of a new neutral gauge boson to the origin of neutrino mass~\cite{Basso:2008iv}. The resulting $Z'_{B-L}$ couples vectorially and flavor-universally, with a charge of $-1$ for leptons and $+1/3$ for quarks. It couples to both the beam electron and the quark, making it a natural resonance to look for in electron-proton collisions. We work in the pure $B-L$ limit and set the gauge-kinetic mixing with hypercharge to zero. This mixing is induced generically  at the loop level~\cite{Holdom:1985ag}. For a boson of a few tens of GeV, the coupling $g_{B-L}$ is already constrained by LEP and LHC data~\cite{Langacker:2008yv, Bauer:2018onh}, and the clean leptonic channel $Z'\to e^+e^-$ is well covered by conventional resonance searches. We therefore consider the hadronic channel, in which the resonance must be reconstructed from jets and separated from a far larger deep-inelastic background. 

\begin{figure}[tp]
 \centering
 \includegraphics[width=.44\linewidth]{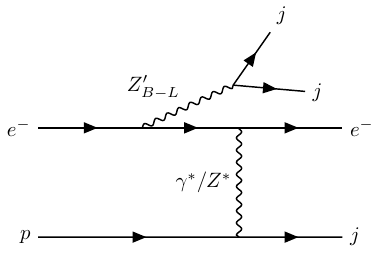}\hfill
 \includegraphics[width=.44\linewidth]{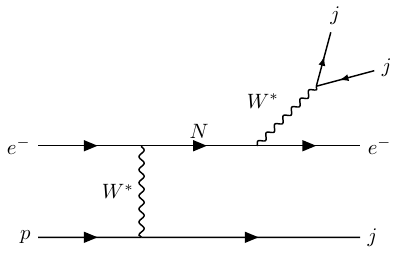}
 \par\smallskip
 \makebox[.44\linewidth]{(a)}\hfill\makebox[.44\linewidth]{(b)}
 \caption{Representative production and decay topologies for (a) the $B-L$ gauge boson and (b) the heavy neutral lepton. Both yield an electron and three hard partons. The electron accompanies resonance production in (a) and belongs to the resonance decay in (b).}
 \label{fig:production}
\end{figure}

We consider production in electron-proton scattering followed by hadronic decay,
\begin{equation}
 e^-p\longrightarrow e^-jZ'_{B-L},\qquad Z'_{B-L}\longrightarrow jj,
 \qquad m_{Z'}=40~\GeV.
 \label{eq:bl-process}
\end{equation}
Fig.~\ref{fig:production}(a) shows the electron-line radiation topology. The observed electron recoils against the resonance, correlating its momentum with the hadronic decay products. Ordering the reconstructed jets by decreasing transverse momentum, we take the invariant mass of the two leading jets as the resonance coordinate,
\begin{equation}
 \mjj^2=(p_{j_1}+p_{j_2})^2.
 \label{eq:m12}
\end{equation}

Neutrino masses may require physics beyond the Standard Model, and one of the simplest modifications to the Standard Model is a fermion with no gauge charges. This state mixes with active neutrinos and, in the type-I seesaw, generates light neutrino masses suppressed by its own mass scale~\cite{Minkowski:1977sc,Mohapatra:1979ia}. As such, a heavy neutral lepton near the electroweak scale is among the best-motivated extensions of the Standard Model. At an electron-proton collider, the mixing $U_{eN}$ creates a unique production mode: the incoming electron converts into the HNL by charged-current scattering off the proton, so the HNL is produced at a rate proportional to $|U_{eN}|^2$. The reach of the EIC in this channel was studied in Ref.~\cite{Batell:2022ogj}, which found sensitivity across the mass range $1$--$100$~GeV, reaching mixing angles of $10^{-4}$--$10^{-3}$ for prompt decays and $10^{-6}$--$10^{-4}$ for displaced ones. Our benchmark lies in the prompt region. When $m_N<m_Z$, the mixing is further constrained by LEP searches for neutral heavy leptons produced in $Z$ decays~\cite{DELPHI:1996qcc}. 

We take $N$ to be a Dirac fermion and study its lepton-number-conserving semileptonic decay. Although the Majorana case is typically more general, the Dirac-like case can be realized in many viable scenarios. For example, a heavy state built from nearly degenerate Majorana components reproduces lepton-number-conserving, Dirac-like behavior, with the residual violation controlled by the mass splitting and mixing~\cite{Gluza:2016qqv}. The observed electron is a decay product of the resonance rather than the scattered beam electron. As such, the resonance must be reconstructed from the lepton together with jets, and the standard deep-inelastic description of the event no longer applies.

We study charged-current production and the semileptonic decay
\begin{equation}
 e^-p\longrightarrow Nj,\qquad N\longrightarrow e^-jj,
 \qquad m_N=50~\GeV.
 \label{eq:hnl-process}
\end{equation}
Here, the identified electron originates from the HNL decay, so the resonance coordinate must combine it with the decay jets. We examine several candidate observables, including invariant masses of the electron with different jet pairs, for example $M_{e j_1 j_2}$. In these samples, the electron combined with the second and third jets creates a resonance peak above the Standard Model background. We use
\begin{equation}
 \mhnl^2=(p_e+p_{j_2}+p_{j_3})^2.
 \label{eq:mhnl}
\end{equation}
We use these fixed jet assignments throughout. They give experimentally accessible resonance coordinates whose distributions include the effects of additional radiation and jet ordering.

We simulate collisions of 18~GeV electrons with 275~GeV protons, corresponding to $\sqrt{s}\simeq141$~GeV, using \textsc{MadGraph5\_aMC@NLO}~\cite{Alwall:2014hca} interfaced to \PYTHIA~8 for parton showering and hadronization~\cite{Sjostrand:2014zea}. The generation setup uses the NNPDF2.3 LO parton distributions~\cite{Ball:2012cx} through LHAPDF~\cite{Buckley:2014ana}. Before showering and hadronization, the parton-level generation settings specify $p_{T,j/e}>1~\mathrm{GeV}$ and $|\eta_{j/e}^{\rm lab}|<4.5$ for the $B-L$ samples, and $p_{T,j/e}>5~\mathrm{GeV}$ and $|\eta_{j/e}^{\rm lab}|<4$ for the HNL samples. Both benchmarks use separation cuts $\Delta R_{jj}>0.4$ and $\Delta R_{je}>0.4$, where $j$ denotes a hard-process quark or gluon. The angular cuts are evaluated in the laboratory frame. These generation cuts are distinct from the particle-level acceptance and reconstruction requirements described below. The Standard Model background consists of deep-inelastic electron-proton events, with QCD radiation contributing to the reconstructed multijet final state. We carry out the analysis at the particle level. We use the HeavyN and minimal $B-L$ models implemented in FeynRules~\cite{Alloul:2013bka,Degrande:2016aje,Basso:2008iv}.

\subsection{Acceptance and resonance reconstruction}
\label{sec:acceptance}

We take the proton beam to point along $+z$ and retain visible final-state particles within the laboratory acceptance
\begin{equation}
 |\eta_{\rm lab}|<4,\qquad
 \eta_{\rm lab}=\frac12\ln\frac{|\boldsymbol p_{\rm lab}|+p_{z,\rm lab}}{|\boldsymbol p_{\rm lab}|-p_{z,\rm lab}}.
 \label{eq:acceptance}
\end{equation}
We exclude neutrinos and require the identified electron to lie inside the acceptance. In the $B-L$ sample, we select the highest-$p_T$ final-state electron. For HNL events, we use the electron candidate recorded in the benchmark sample.

The stored particle coordinates are in the electron-proton center-of-mass frame. We boost the four-momenta to the laboratory frame to apply the acceptance cuts, then use the retained collection of momenta in the center-of-mass frame for clustering and analysis. Neglecting the beam masses, the boost is specified by
\begin{equation}
 \begin{aligned}
 y_{\rm lab}&=y_{\rm cm}+Y,
 &Y&=\frac12\ln\frac{E_p}{E_e}\simeq1.363,\\
 y_{\rm cm}&=\frac12\ln\frac{E_{\rm cm}+p_{z,\rm cm}}{E_{\rm cm}-p_{z,\rm cm}}.
 \end{aligned}
 \label{eq:boost}
\end{equation}
Here, $E_e$ and $E_p$ are the laboratory beam energies, and $Y$ is the boost rapidity. Because rapidity $y_{\rm cm}$ and pseudorapidity differ for a massive particle, equation~\eqref{eq:boost} is not used to impose the acceptance directly: each particle's four-momentum is boosted with its mass retained, and $\eta_{\rm lab}$ is evaluated from the resulting three-momentum. For the rest of this work, rapidities and energies do not carry any frame subscript for simplicity and always refer to the center-of-mass frame unless stated otherwise.

We remove the identified electron from the clustering inputs and reconstruct jets from the remaining accepted particles using the anti-$k_T$ algorithm~\cite{Cacciari:2008gp} with E-scheme recombination in \textsc{FastJet}~\cite{Cacciari:2011ma}. The resonance coordinates in Eqs.~\eqref{eq:m12} and~\eqref{eq:mhnl} are calculated from these jets. We require at least two jets for the $B-L$ candidate and three for the HNL candidate.

\begin{figure}[tp]
 \centering
 \includegraphics[width=\linewidth]{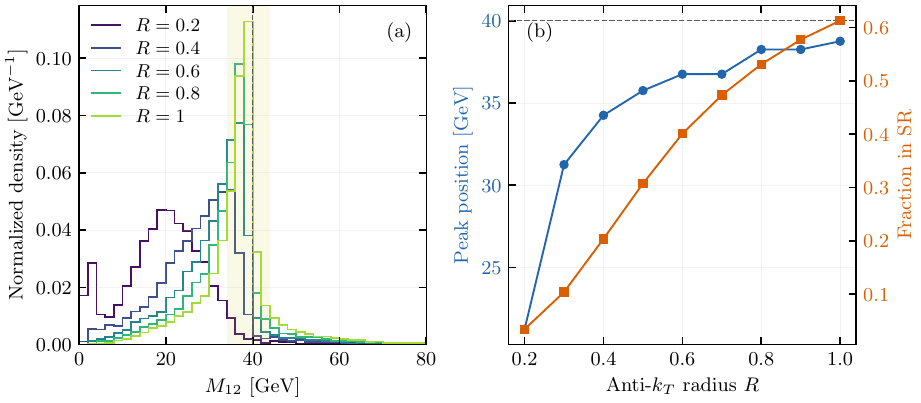}
 \caption{Jet-radius dependence of the $B-L$ resonance reconstruction with laboratory acceptance. Left: normalized leading-dijet mass distributions for representative radii. Right: peak position (blue), defined as the center of the highest-count bin in $20<\mjj<60$~GeV, and fraction in the signal region (orange). The dashed line marks the generated mass, and the shaded interval is the signal region. We use a common sample of 10,000 signal events. For each radius, we normalize to events with an accepted electron and a valid candidate in $0<\mjj<80$~GeV.}
 \label{fig:radius}
\end{figure}

Figure~\ref{fig:radius} shows how the jet radius affects the reconstruction of the $B-L$ resonance. Narrow jets miss part of the hadronic decay products, moving the peak below the generated mass. Widening the radius recaptures them and raises the fraction of signal events in the nominal signal region. The peak position in the right panel is the center of the most populated mass bin in the search window. The large radius favored here follows from the modest boost of the resonance: at $m_{Z'}/\sqrt{s}\simeq 0.3$ its decay products are widely separated in angle. Similar considerations apply to HNL reconstruction: radiation outside the candidate decay jets can lower $\mhnl$, while the jet ordering affects which particles enter the candidate. We adopt $R=1$ for both benchmarks.

We write $\mres$ when referring to the resonance coordinate of either benchmark. Our mass windows and signal regions (SRs) are
\begin{equation}
\begin{array}{c|cc}
 & \text{mass window [GeV]} & \text{signal region [GeV]}\\ \hline
B-L & 20<\mjj<60 & 34<\mjj<44\\
\text{HNL} & 40<\mhnl<70 & 43<\mhnl<53.
\end{array}
\label{eq:windows}
\end{equation}
The signal region (SR) contains the hypothesized resonance peak, while the sidebands (SBs) are the background-dominated parts of the mass window on either side. For an unknown resonance mass, the procedure can be repeated over a sliding mass window~\cite{Collins:2018epr}, with the significance corrected for the look-elsewhere effect. The fixed windows above define the benchmarks studied here.

\subsection{Event representation and observables}
\label{sec:observables}

We retain up to 150 final-state particles per event and mask any padding. The full-event classifier receives the accepted non-electron particles, including those outside the leading jets, together with event-level information about the electron. This representation retains correlations between the electron, reconstructed jets, and particles not assigned to those jets. 

Several high-level observables help interpret the separation between signal and background. We define the visible transverse activity and transverse-momentum imbalance as
\begin{equation}
 H_T^{\rm vis}=\sum_{i\in\mathcal V}p_{Ti},\qquad
 p_{T,\rm miss}^{\rm vis}=\left|\sum_{i\in\mathcal V}\boldsymbol p_{Ti}\right|,
 \label{eq:ht}
\end{equation}
where $\mathcal V$ contains all accepted visible particles, including the electron. The momentum imbalance receives contributions from particles outside the acceptance as well as invisible particles. The particle multiplicity $N$ denotes the size of the non-electron collection.

For incident electron, observed electron, and incident proton momenta $k$, $k'$, and $P$, respectively, the electron-method DIS variables are
\begin{equation}
 q=k-k',\qquad Q^2=-q^2,\qquad
 x_B=\frac{Q^2}{2P\cdot q},\qquad
 y_{\rm DIS}=\frac{P\cdot q}{P\cdot k}.
 \label{eq:dis}
\end{equation}
Here $y_{\rm DIS}$ is the DIS inelasticity. For the $B-L$ benchmark, $x_B$ and $Q^2$ are informative precisely because radiation of the $Z'_{B-L}$ distorts their relation to the underlying hard scattering. The HNL electron is instead a decay product, so in that classifier we replace these two inputs by $H_T^{\rm vis}$ and $p_{T,\rm miss}^{\rm vis}$.

For each particle, we define the dimensionless energy variable
\begin{equation}
 z_i=\frac{2E_i}{\sqrt{s}},\qquad
 E_i=(p_{Ti}^2+m_i^2)^{1/2}\cosh y_i.
 \label{eq:z}
\end{equation}
The particle masses follow the recorded species. We retain these masses in the energy variable, following Ref.~\cite{Araz:2024bom}. For a complete final state in the center-of-mass frame, energy conservation gives $\sum_{i\in\mathrm{event}}z_i=2$. The finite laboratory acceptance means that this sum need not equal two. The full-event classifier nevertheless retains the electron kinematics through its event-level inputs and through the electron-relative particle coordinates, as detailed in section~\ref{sec:learning}. 

%As a compact summary of how the hadronic energy is distributed relative to the electron, we use the energy-weighted mean rapidity displacement,
%\begin{equation}
%\label{eq:ydisp}
% \langle y-y_e\rangle_E
% =\frac{\sum_i E_i(y_i-y_e)}{\sum_i E_i}
% =\frac{\sum_i z_i(y_i-y_e)}{\sum_i z_i},
%\end{equation}
%where the sums run over accepted non-electron particles. In the language of energy-flow observables~\cite{Chen:2020vvp}, this quantity is the normalized first rapidity moment of the accepted hadronic energy flow and indicates whether its energy-weighted center lies forward or backward of the electron. Related energy-flow observables have been studied in deep-inelastic scattering, and for the EIC~\cite{Liu:2022wop,Fu:2025hpc,Kang:2026hig}, while transverse energy-flow distributions have been measured at HERA~\cite{H1:1999dbb}. 

As a compact summary of how the hadronic energy is distributed relative to the electron, we calculate the energy-weighted mean rapidity displacement separately for each event,
\begin{equation}
\label{eq:ydisp}
 \langle y-y_e\rangle_E
 =\frac{\sum_i E_i(y_i-y_e)}{\sum_i E_i}
 =\frac{\sum_i z_i(y_i-y_e)}{\sum_i z_i},
\end{equation}
where the sums run over accepted non-electron particles in that event. The energies and rapidities are evaluated in the COM frame. This quantity is the normalized first rapidity moment of the  energy flow~\cite{Chen:2020vvp} and indicates whether its energy-weighted center lies forward or backward of the electron. Related energy-flow observables have been studied in deep-inelastic scattering and for the EIC~\cite{Liu:2022wop,Fu:2025hpc,Kang:2026hig}, while transverse energy-flow distributions have been measured at HERA~\cite{H1:1999dbb}. For the boosted decision tree, we use the first four energy-weighted moments of the rapidity displacement,
\begin{equation}
 M_n^{(E)}
 =\frac{\sum_i E_i(y_i-y_e)^n}{\sum_i E_i}
 =\frac{\sum_i z_i(y_i-y_e)^n}{\sum_i z_i},
 \qquad n=1,2,3,4,
 \label{eq:bdt-energy-moments}
\end{equation}
with $M_1^{(E)}=\langle y-y_e\rangle_E$. Each moment is computed within a single event and supplied as a separate classifier input.

In the topology plots, angular separations are $\Delta R_{ab}=[(\Delta\eta_{ab})^2+(\Delta\phi_{ab})^2]^{1/2}$, where $\Delta\eta_{ab}=\eta_a-\eta_b$ and $\Delta\phi_{ab}=\phi_a-\phi_b$ wrapped to $(-\pi,\pi]$. These use pseudorapidity, whereas Eq.~\eqref{eq:ydisp} uses true rapidity relative to the electron. For the HNL case, we also use the minimum separation between the electron and a candidate decay jet, $\min\{\Delta R_{e j_2},\Delta R_{e j_3}\}$.  Together, these summary statistics describe both the overall hardness and the relative arrangement of the particles.

\section{Weakly supervised classification}
\label{sec:learning}

\subsection{Learning from mixed event samples}

\begin{figure}[tp]
 \centering
 \includegraphics[width=\linewidth]{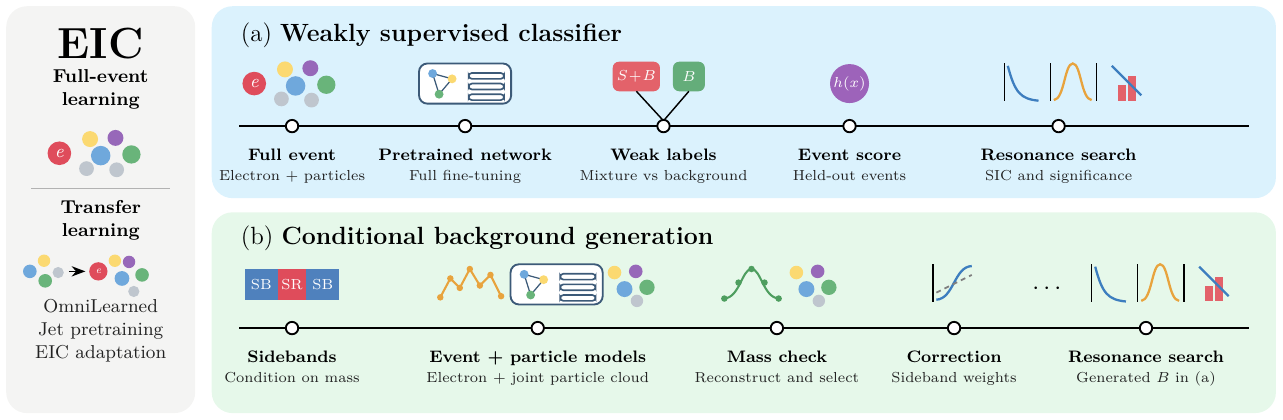}
 \caption{Full-event learning for the EIC. The weakly supervised classifier combines the particle collection and electron kinematics into an event score. The conditional generator in (b) learns event conditions and particle distributions from background sidebands, followed by reconstructed-mass selection and a sideband-derived probability correction. Its output supplies the background reference $B$ for the classifier and resonance-search chain in (a). Both applications adapt pretrained particle-model weights~\cite{Bhimji:2025isp}. The generator's auxiliary event-condition networks are trained separately on EIC events.}
 \label{fig:workflow}
\end{figure}

We employ weakly supervised anomaly detection~\cite{Collins:2018epr}, based on the classification without labels (CWoLa) paradigm~\cite{Metodiev:2017vrx}. The training task is to distinguish a sample containing signal plus background, $S+B$, from an independent background sample, $B$. Each event is labeled only by which sample it belongs to, instead of being labeled as a signal or background event. Figure~\ref{fig:workflow} summarizes the analysis workflow and shows how a generator trained on background sidebands supplies the background sample $B$ for the classifier and resonance-search chain. For a signal fraction $f_S$ in the SR, the positive training class follows the mixture
\begin{equation}
 p_{\rm mix}(x)=(1-f_S)b(x)+f_Ss(x),\qquad
 f_S=\frac{S}{S+B},\qquad \frac{S}{B}=\frac{f_S}{1-f_S},
 \label{eq:mixture}
\end{equation}
where $S$ and $B$ are the expected signal and background counts in the mixed sample. The vector $x$ contains the measured inputs for one event, while $s(x)$ and $b(x)$ are their probability densities in pure signal and pure background, each normalized to one. They describe how frequently different event configurations occur. The network learns a score $h(x)$ for distinguishing the two training samples. With identical background distributions in the samples, an ideal classifier ranks events by the mixed-to-background density ratio,
\begin{equation}
 \frac{p_{\rm mix}(x)}{b(x)}=1+f_S\left[\frac{s(x)}{b(x)}-1\right].
 \label{eq:cwola}
\end{equation}
We denote by $D(x)$ the probability assigned to the mixed training class. For balanced training classes, an ideal classifier gives
\begin{equation}
     D(x)=\frac{p_{\rm mix}(x)}{p_{\rm mix}(x)+b(x)}.
\end{equation} 
Its odds $D(x)/[1-D(x)]$ equal the ratio in equation~\eqref{eq:cwola}. This probability refers to the training-sample label, not to whether an individual event is signal. For $f_S>0$, the ratio increases with $s(x)/b(x)$, so ranking by the learned score can also enrich signal. The Neyman-Pearson lemma therefore connects this weakly supervised objective to optimal signal selection~\cite{Neyman:1933wgr,Metodiev:2017vrx}.

In the main analysis, the reference background is drawn from the same simulated distribution as the background in the mixed sample. This corresponds to the limit of an ideal generative model and isolates the information that the classifier can extract. In section~\ref{sec:generator}, we replace this reference by events generated from a sideband-trained model, where the task of this generative model is to create realistic background events inside the SR. Any mismatch between the generated and true background can affect the downstream classification task, potentially reducing the signal enhancement.

We match the two training classes in 0.25~GeV bins of reconstructed resonance mass within the SR. This prevents the injected mass peak alone from distinguishing the classes. The mass itself is omitted from the inputs. Absolute particle momenta still carry mass information, so we also control the mass dependence of the final selection, as described in section~\ref{sec:inference}. Without this control, the classifier could distinguish the training samples through their resonance-mass distributions and sculpt the selected background spectrum~\cite{Geirhos:2020shortcut,Benkendorfer:2020gek,Nachman:2019dol}.

We adopt a five-fold cross-validation. For each test fold, a different fold is used for validation, and the remaining three for training. Each fit uses 115,000 training and 30,000 validation events per weak class. We select checkpoints based on the minimum weak-label validation loss and combine predictions on the five held-out folds. Repeating this procedure for three random seeds gives an estimate of training variability. Each signal fraction requires separate training. Signal labels are used only to evaluate trained classifiers.

\subsection{Full-event particle classifiers}

Our main classifier is an OmniLearned point-edge transformer, PET2~\cite{Bhimji:2025isp}, with 2.76 million parameters after adapting it to accommodate EIC events. Its pretraining on jets provides a starting point for learning correlations among particles. We adapt the input representation and the classification head to complete EIC events and fine-tune all weights. This extends the use of particle-physics foundation models across processes and tasks~\cite{Mikuni:2024qsr,Birk:2024knn}.

For each accepted non-electron particle, we construct
\begin{equation}
 x_i=\left(\log_{10}\frac{p_{Ti}}{p_{Te}},\ \Delta y_{ie},\
 \Delta\phi_{ie},\ \log_{10}z_i,\ Q_i^{\rm em}\right),
 \label{eq:particle-inputs}
\end{equation}
where $Q_i^{\rm em}$ is the electric charge in units of the positron charge, and we have $\Delta y_{ie}=y_i-y_e$, $\Delta\phi_{ie}=\phi_i-\phi_e$ are the polar and azimuthal angles relative to the electron, respectively. We embed the absolute PDG identifier as a categorical feature and provide electric charge separately, so particles and antiparticles of the same species share an embedding. The event-level inputs are
\begin{align}
{\bf u}_{B-L}&=\left(\log_{10}p_{Te},\ y_e,\ \sin\phi_e,\ \cos\phi_e,
 \log_{10}x_B,\ \log_{10}Q^2\right),\label{eq:globals-bl}\\
 {\bf u}_{\rm HNL}&=\left(\log_{10}p_{Te},\ y_e,\ \sin\phi_e,\ \cos\phi_e,
 \log_{10}H_T^{\rm vis},\ \log_{10}p_{T,\rm miss}^{\rm vis}\right).
 \label{eq:globals-hnl}
\end{align}
Small positive floors make the logarithms finite. The inputs differ because the $B-L$ electron accompanies the scattering, while the HNL electron comes from a decay. We use DIS kinematics in the first case and measures of visible event activity in the second. We retain the electron azimuth so that the representation can also accommodate polarized-scattering applications at the EIC~\cite{Wang:2024zns}.

The input adapter converts these variables to the angular and logarithmic momentum coordinates used by PET2. We replace the jet-level conditioning with the six event-level inputs in equation~\eqref{eq:globals-bl} or~\eqref{eq:globals-hnl}, together with $\ln(1+N)$, where $N$ is the number of accepted non-electron particles defined in section~\ref{sec:observables}. New embeddings accommodate the EIC particle categories. Local particle interactions and attention across the event then build a representation that includes particles inside and outside the reconstructed jets. The classification head combines this representation into a single event score. In particular, we feed the entire particle collection to one network, allowing it to exploit correlations among the jets, the electron, and accepted particles outside those jets.

Approximately 20,000 parameters are newly initialized in the input adaptation, conditioning, and output components. We adopt the Lion optimizer~\cite{Chen:2023lion}, with learning rates $5\times10^{-6}$ for transferred weights and $2.5\times10^{-5}$ for new parameters. Training uses weight decay and dropout of 0.1, one warm-up epoch followed by cosine decay, and early stopping with a patience of eight epochs within a maximum of 40 epochs.

We also explored other classifiers, including a Particle Flow Network (PFN)~\cite{Komiske:2018cqr,Zaheer:2017ue} and a Particle Transformer (ParT)~\cite{Qu:2022mxj}. Their performance was broadly comparable or lower for the data sets used here. A systematic quantitative comparison with matched tuning and varied training-sample sizes is left to future work.

\subsection{Summary statistics}

To understand how much of the full-event sensitivity can be captured by summary statistics, we compare the PET2 classifier with two high-level classifiers for the $B-L$ benchmark. One uses only the variables $(x_B,Q^2)$, which describe the scattered electron. The other takes 36 inputs that summarize the electron kinematics and the accepted non-electron particle collection. Figure~\ref{fig:ladder-extra} compares these two classifiers with the full-event classifier and the mass-only search.

The 36-input representation contains the six electron/DIS quantities in equation~\eqref{eq:globals-bl}, with the transformations specified there, and 30 statistics of the accepted non-electron particles. These particle statistics comprise the multiplicity $N$, 16 kinematic summaries, three energy and charge sums, and ten species fractions. The 16 kinematic summaries are the mean, standard deviation, minimum, and maximum of each of the first four particle inputs in equation~\eqref{eq:particle-inputs}. Each particle has equal weight in the mean and standard deviation, and the variance is normalized by $N$. The three sums are $\sum_i z_i$, using the energy variable in equation~\eqref{eq:z}, the net charge $\sum_iQ_i^{\rm em}$, and the absolute-charge sum $\sum_i|Q_i^{\rm em}|$.

The species fractions are $f_c=N_c/N$, where $N_c$ counts particles in category $c$. The ten categories are electrons, electron neutrinos, muons, muon neutrinos, tau neutrinos, photons, long-lived neutral kaons, charged pions, charged kaons, and neutrons, with antiparticles merged. The visible-particle selection excludes neutrinos. The three neutrino-fraction entries are retained in the stored 36-input vector but are identically zero and provide no information to the classifier. The electron fraction counts any additional electrons after the identified electron has been removed from the particle collection. The identified electron is represented by the event-level inputs. These ten fractions are part of the stored representation and do not cover all species, so they need not sum to one.

We train boosted decision trees for both the two-input and the 36-input representations using histogram-based gradient boosting in \textsc{scikit-learn}~\cite{scikit-learn}. Both use 100 boosting iterations, a learning rate of 0.08, at most 31 leaves, a minimum leaf population of 20, and $L_2$ regularization of 1.0. Each fit uses 130,000 training and 30,000 validation events per weak class. Figure~\ref{fig:ladder-extra} uses one five-fold training for each classifier.

Particle counts, energy-flow correlations, and radiation patterns have proven useful in jet classification \cite{Frye:2017yrw, Komiske:2017aww, Dreyer:2018nbf, Athanasakos:2023fhq}. At the EIC, their correlations with the electron and the rest of the hadronic event can provide additional information. The two-input classifier tests the sensitivity available from DIS kinematics alone. Comparing the 36-input classifier with the full-event classifier tests how much of the full-event sensitivity is retained by the chosen summary statistics~\cite{Finke:2023ltw}.

\section{From classifier score to significance}
\label{sec:inference}

The classifier provides an event ranking that we use to select signal-enriched samples. To turn this ranking into a resonance search, we need to calibrate its mass dependence and compare the selected mass distribution with a background prediction constrained by the sidebands. 

\subsection{Classification performance}

For a threshold $t$, on the score $h$, the signal and background efficiencies are $\epsS(t)=P_S(h>t)$ and $\epsB(t)=P_B(h>t)$. Here, $P_S$ ($P_B$) is the probability for an event drawn from the signal (background) population. Thus, each efficiency is the fraction of that population passing the score cut $h>t$, estimated by counting passing events in independent signal and background test samples. In terms of the densities in section~\ref{sec:learning}, for example, $P_S(h>t)=\int_{h(x)>t}s(x)\,dx$. The receiver operating characteristic (ROC) curve records their trade-off, and its area (AUC) measures separation over the full range of thresholds. Throughout, the signal-versus-background AUC is evaluated with simulation truth after training. For an exact background reference and a fixed score, the mixed-versus-background AUC satisfies
\begin{equation}
 \mathrm{AUC}_{\rm mix,B}=\frac12+
 f_S\left(\mathrm{AUC}_{S,B}-\frac12\right).
 \label{eq:weakauc}
\end{equation}
Thus, the weak-training AUC can be close to $1/2$ even when the score separates signal and background well.

Search sensitivity is driven by how much signal survives strong background rejection. We quantify this through the significance-improvement characteristic (SIC)~\cite{Gallicchio:2010dq},
\begin{equation}
 \SIC(t)=\frac{\epsS(t)}{\sqrt{\epsB(t)}}.
 \label{eq:sic}
\end{equation}
This is the factor by which the selection improves $S/\sqrt{B}$ in the background-dominated limit. We show both the empirical maximum and a protected maximum defined by requiring at least 100 surviving test-background events. The latter limits fluctuations from the sparsely populated end of the curve. For $N_B^{\rm test}$ test-background events, it probes $\epsB\geq100/N_B^{\rm test}$.

We report means over three training seeds, with the standard error of the mean (SEM) unless stated otherwise. Each caption states whether a band shows the spread of curves or the error on a maximum. We calculate the maximum separately for each seed before averaging. These errors quantify classifier-training variability for the fixed simulation samples.

\subsection{Mass-dependent score calibration}
\label{sec:calibration}

The classifier inputs contain correlations with the resonance mass. Consequently, a fixed score threshold can select some mass intervals more efficiently than others and distort the background spectrum. Several methods have been developed to control this effect in resonant searches~\cite{Kasieczka:2020yyl, Benkendorfer:2020gek, Hallin:2022eoq,Das:2024eie}. We use the background score distribution at each mass to define a percentile,
\begin{equation}
 u=F_B(h\mid\mres),\qquad
 F_B(h\mid m)=P_B(H\leq h\mid\mres=m).
 \label{eq:pit}
\end{equation}
For a continuous score and an exact reference, $u$ is uniform for background at each mass. A cut $u>1-\epsB$ then has a fixed background efficiency across the spectrum. Mass-dependent thresholds also underlie designed decorrelated taggers~\cite{Dolen:2016kst}.

We determine the percentiles in bins of 1~GeV for $B-L$ and 0.5~GeV for HNL and apply the resulting selection to held-out search events. These bin widths specify the calibration and displayed spectra. The finer 0.25~GeV bins introduced in section~\ref{sec:learning} instead match the training classes. We inspect the selected background on independent events to assess residual mass dependence. The reference in this study is available throughout the mass window. 

\subsection{Sideband-constrained likelihood}

After selection, we use an independent background sample to predict the mass shape and the search sidebands to determine its normalization. The sidebands provide a control sample in which little signal is expected.  Once the reference shape is fixed, they set how many background events are predicted in the signal region. The smooth line in our mass-spectrum plots is this normalized reference prediction. We test for a positive excess in the total SR yield, leaving the distribution of that excess inside the SR unspecified.

Let $n_i$ and $r_i$ denote search and reference counts in sideband bin $i$. The corresponding totals in the SR are $N_{\rm SR}$ and $R_{\rm SR}$. We introduce a relative normalization $\alpha$, unknown reference expectations $\beta_i$ and $\beta_{\rm SR}$, and an excess yield $\mu\geq0$. The nuisance parameters $\alpha$ and $\beta$ describe the uncertain background normalization and bin expectations. They are fitted together with the possible excess rather than treated as exactly known. Writing $\mathrm{Pois}(n\mid\lambda)$ for the probability of observing count $n$ when the expected count is $\lambda$, the likelihood is
\begin{align}
 L(\mu,\alpha,\{\beta_i\},\beta_{\rm SR})={}&
 \prod_{i\in\mathrm{SB}}
 \mathrm{Pois}(n_i\mid\alpha\beta_i)\,
 \mathrm{Pois}(r_i\mid\beta_i)\nonumber\\[-2pt]
 &\times\mathrm{Pois}(N_{\rm SR}\mid\alpha\beta_{\rm SR}+\mu)\,
 \mathrm{Pois}(R_{\rm SR}\mid\beta_{\rm SR}).
 \label{eq:likelihood}
\end{align}
The likelihood combines the observed counts in the search and reference samples. Its first line describes the background-dominated sidebands. Its second line permits an additional yield $\mu$ only in the search signal region. The sideband terms constrain $\alpha$. The reference SR count determines the transfer into the signal region, while profiling the $\beta$ parameters accounts for finite reference statistics. With one common normalization and an aggregate SR excess, the sideband and SR totals contain the information needed for this test. The displayed finer binning lets us inspect the shape of the selected spectrum.

Following Ref.~\cite{Cowan:2010js}, we use the one-sided profile-likelihood statistic
\begin{equation}
q_0=-2\ln\frac{L(0,\widehat{\widehat{\boldsymbol\theta}})}
 {L(\widehat\mu,\widehat{\boldsymbol\theta})},\qquad
 Z_{\rm local}\simeq \sqrt{q_0},
 \label{eq:q0}
\end{equation}
where $\boldsymbol\theta$ collects the nuisance parameters. Single hats denote the values that maximize the likelihood when the excess and nuisance parameters are fitted together. The double-hatted values maximize it subject to the background-only hypothesis $\mu=0$. The ratio, therefore, compares the best background-only description with the best description allowing for an excess. A lower ratio, and hence a larger $q_0$, indicates a stronger preference for a positive excess. We set $q_0=0$ if the unconstrained best fit favors a negative signal yield. The conversion $Z_{\rm local}\simeq\sqrt{q_0}$ uses the asymptotic approximation of Ref.~\cite{Cowan:2010js}.

To estimate the expected sensitivity, we use the background counts predicted by the reference sample after normalizing it in the sidebands and add the expected signal counts. This Asimov construction estimates the typical sensitivity without relying on an upward or downward fluctuation of the search sample. We also check the background prediction against an independent background sample, which we use in the displayed mass spectra.

For each benchmark, we use a signal fraction $f_S=0.01$ and choose the total event yield such that the expected local significance is $Z_{\rm local}\approx 1$. This yield is held fixed across classifier selections, so changes in $Z_{\rm local}$ quantify the selection gain at fixed statistics.

We evaluate three predefined background efficiencies. The quoted significance applies to the fixed benchmark mass window. A search over unknown masses would also need to account for the possibility of finding a background fluctuation somewhere in the scanned range~\cite{Choudalakis:2011qn}. Training and evaluation events are kept separate to prevent the classifier from exploiting statistical fluctuations in its training sample~\cite{Hein:2025ysv}. Converting these significances into luminosity or coupling reach would require physical cross sections, detector efficiencies, and systematic uncertainties, which we leave for future work.

\section{Results}
\label{sec:results}

\subsection{Kinematic origin of the \texorpdfstring{$B-L$}{B-L} separation}

The separation between the $B-L$ signal and the Standard Model background is primarily kinematic. Both contain an electron and several resolved jets. The signal therefore differs from many boosted-particle searches at the LHC, where the substructure of a multi-prong decay can be compared with a predominantly single-prong QCD jet. Here the electron recoils against a massive object, changing its kinematics and the distribution of the accompanying hadronic activity.

\begin{figure}[tp]
 \centering
 \includegraphics[width=.98\linewidth]{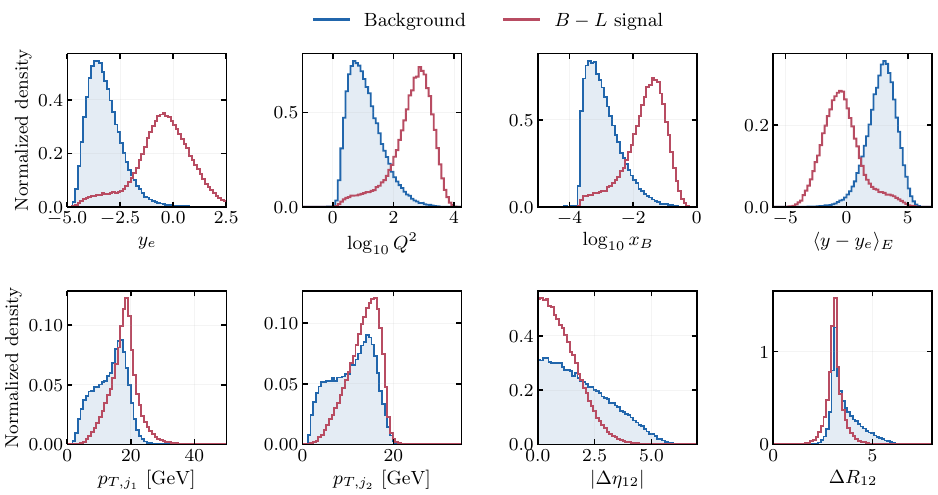}
 \caption{Normalized $B-L$ signal and background distributions after $|\eta_{\rm lab}|<4$ acceptance. We match the reconstructed SR mass in 0.25~GeV bins, using 100,000 events per class. The observables are defined in section~\ref{sec:observables}. The differences shown here provide information beyond the marginal dijet-mass distribution.}
 \label{fig:bl-physics}
\end{figure}

Figure~\ref{fig:bl-physics} illustrates these differences after matching the signal and background in reconstructed mass within the SR. The scattered electron is more central in signal events, and the distributions of $x_B$ and $Q^2$ shift accordingly. The jet momenta and their angular separation also differ. For a given dijet mass, the continuum background can populate configurations with softer jets at larger separation, whereas the resonance decay favors a harder, more compact pair. The remaining event activity carries information about how this pair is produced and how it recoils against the electron and the third jet.

\begin{figure}[tp]
 \centering
 \includegraphics[width=.98\linewidth]{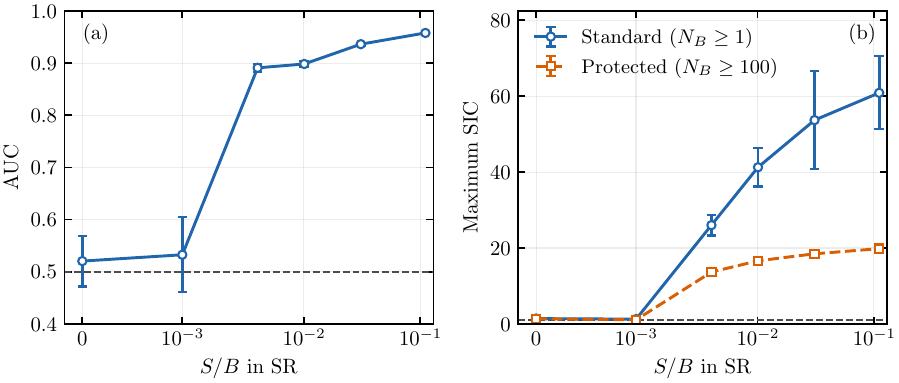}
 \caption{$B-L$ classifier performance versus $S/B=f_S/(1-f_S)$ with laboratory acceptance. Left: signal-versus-background AUC. Right: maximum SIC, evaluated over all thresholds with at least one surviving background event and over the restricted range with at least 100. Points are means over three five-fold trainings, and error bars show SEM. The scan contains $f_S=0$, 0.001, 0.004, 0.01, 0.03, and 0.10.}
 \label{fig:bl-fractions}
\end{figure}

We first ask how much of this information can be learned from a small signal admixture. Figure~\ref{fig:bl-fractions} shows the classifier performance as a function of the injected ratio $S/B$. In the left panel, the AUC remains close to chance-level classification below $f_S=0.001$, then increases to approximately 0.9 at $f_S=0.004$ and improves further at larger fractions. The right panel shows substantial signal enrichment, but also a sizable difference between the two definitions of maximum SIC. At $f_S=0.01$, the mean maximum is approximately 41 when allowing thresholds with at least one surviving test-background event, compared to 17 when requiring at least 100. The larger maxima occur at very tight selections where only a few background events survive, making the estimated background efficiency extremely sensitive to statistical fluctuations. Requiring at least 100 events excludes these selections and gives lower maxima with much smaller variation between training runs. Both curves use the same trained classifiers. The signal fraction at which weak supervision becomes effective depends on the number of training events and the information available to the network.

The comparisons below focus on the expected local significance after background rejection. They assess the information in the event representation at the search working points, without assigning a general ranking to different classifier architectures.

\begin{figure}[tp]
 \centering
 \includegraphics[width=.72\linewidth]{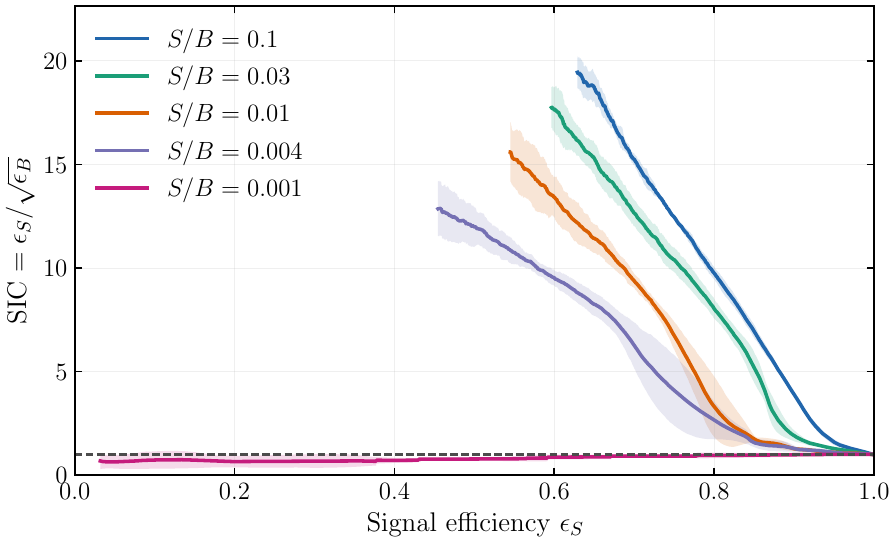}
 \caption{$B-L$ significance-improvement curves for the nonzero injected fractions in figure~\ref{fig:bl-fractions}, with laboratory acceptance and PID enabled. Lines are pointwise means over three seeds after five-fold raw-score pooling, and bands show one standard deviation. Each seed is evaluated on $10^5$ signal and $10^5$ background events. Each curve ends where fewer than 100 test-background events survive in any seed. Legend values give $S/B=f_S/(1-f_S)$.}
 \label{fig:bl-sic}
\end{figure}

Figure~\ref{fig:bl-sic} shows the full significance-improvement curves for the nonzero fractions in figure~\ref{fig:bl-fractions}. All curves use the main classifier with PID enabled. The classifier achieves useful signal enrichment over a broad range of signal efficiencies once the injected fraction is large enough.

\subsection{Resonance search with the full-event classifier}

\begin{figure}[tp]
 \centering
 \includegraphics[width=\searchFigureWidth,trim=0 20bp 0 0,clip]{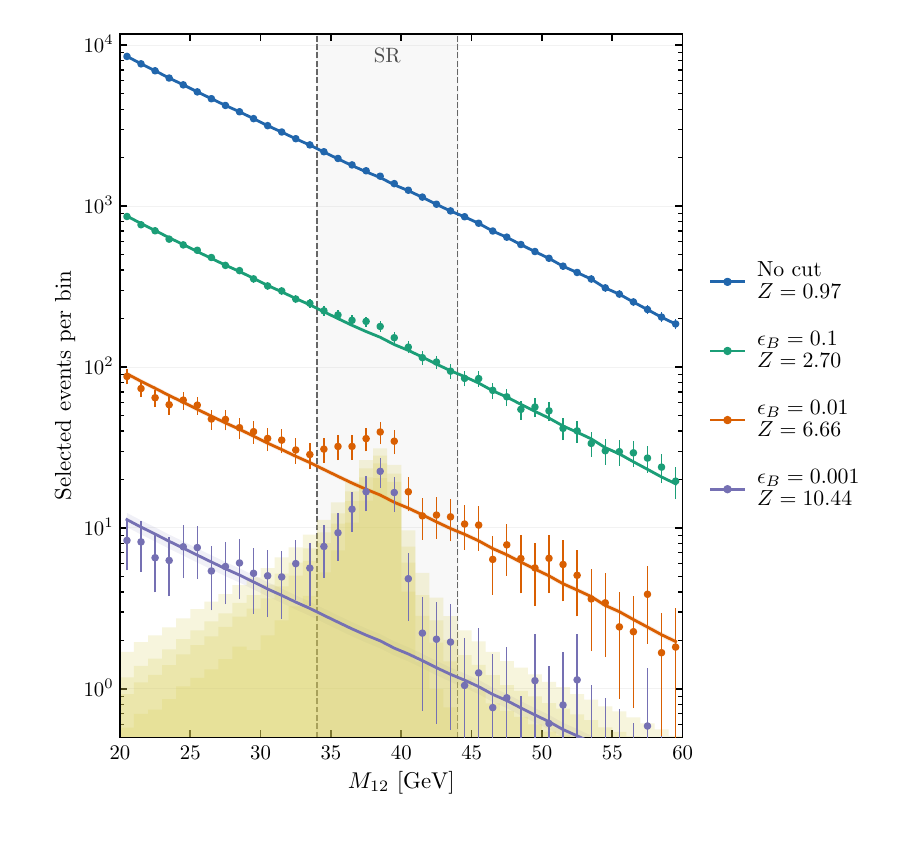}
 \caption{$B-L$ mass spectra before and after classifier selection at $f_S=0.01$ with laboratory acceptance. Points show an independent search-background realization plus the injected signal. Continuous curves give the sideband-normalized background prediction throughout the mass window, and yellow histograms show the signal contribution. Vertical lines and shading delimit the SR. All selections use the same exposure and one five-fold classifier training. The quoted $Z$ values are expected local significances evaluated with the finite-reference likelihood.}
 \label{fig:bl-bump}
\end{figure}

Figure~\ref{fig:bl-bump} shows the main search result: the dijet-mass spectra of the mixed sample after selection on the classifier score. We calibrate the score in mass bins, normalize the reference in the sidebands, and evaluate the expected local significance with the likelihood of section~\ref{sec:inference}. For this benchmark, full-event weakly supervised classification achieves an order-of-magnitude gain in expected local significance, from $0.97$ before selection to $10.44$ at the tightest working point, with the same exposure. This gain demonstrates the method's ability to reveal a resonance that would remain statistically insignificant in the inclusive mass spectrum.

\begin{figure}[tp]
 \centering
 \includegraphics[width=.72\linewidth]{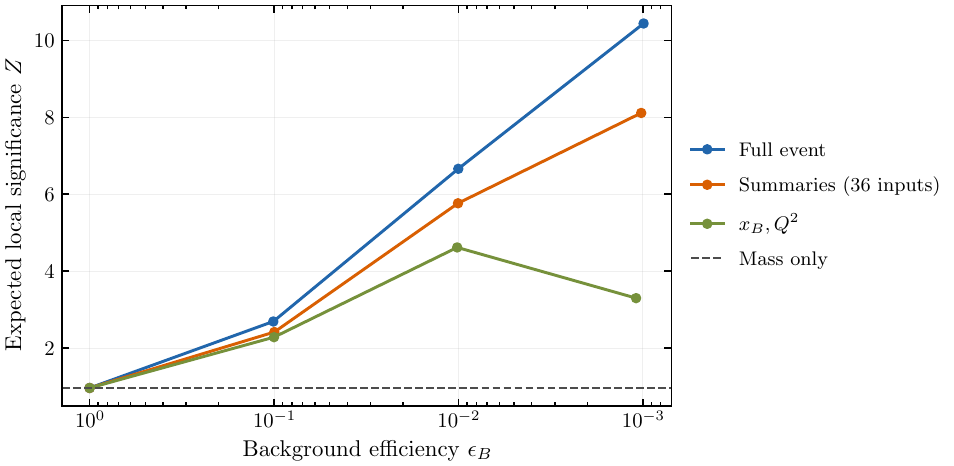}
 \caption{Expected local significance for the $B-L$ benchmark at $f_S=0.01$ for classifiers using $(x_B,Q^2)$, the 36 summary inputs, or the full event. All curves use the same finite-reference likelihood and exposure. Each classifier uses one five-fold training. The dashed line is the mass-only search. The summaries contain all six electron/DIS inputs plus the particle statistics defined in section~\ref{sec:learning}.}
 \label{fig:ladder-extra}
\end{figure}

The comparison with summary statistics is shown in figure~\ref{fig:ladder-extra}. The two DIS variables improve the mass-only search. The 36-input summaries include all six electron/DIS quantities in equation~\eqref{eq:globals-bl} plus the additional particle statistics defined in section~\ref{sec:learning} and give a stronger compact baseline. The full-event classifier gives the largest significance at each of the three working points. Its advantage grows toward tighter selections in this comparison. The full-event classifier therefore provides additional sensitivity beyond these summaries in the part of phase space most relevant to the search.

\subsection{Heavy neutral lepton benchmark}

We next test the same approach when the electron plays a different role. In the HNL process, $ep\to Nj$, the final-state electron is produced in the decay of the heavy neutral lepton. It is therefore typically harder and more central than a background DIS electron. The DIS variables inferred from this electron no longer describe the production interaction, which motivates the alternative global inputs of equation~\eqref{eq:globals-hnl}.

\begin{figure}[tp]
 \centering
 \includegraphics[width=.98\linewidth]{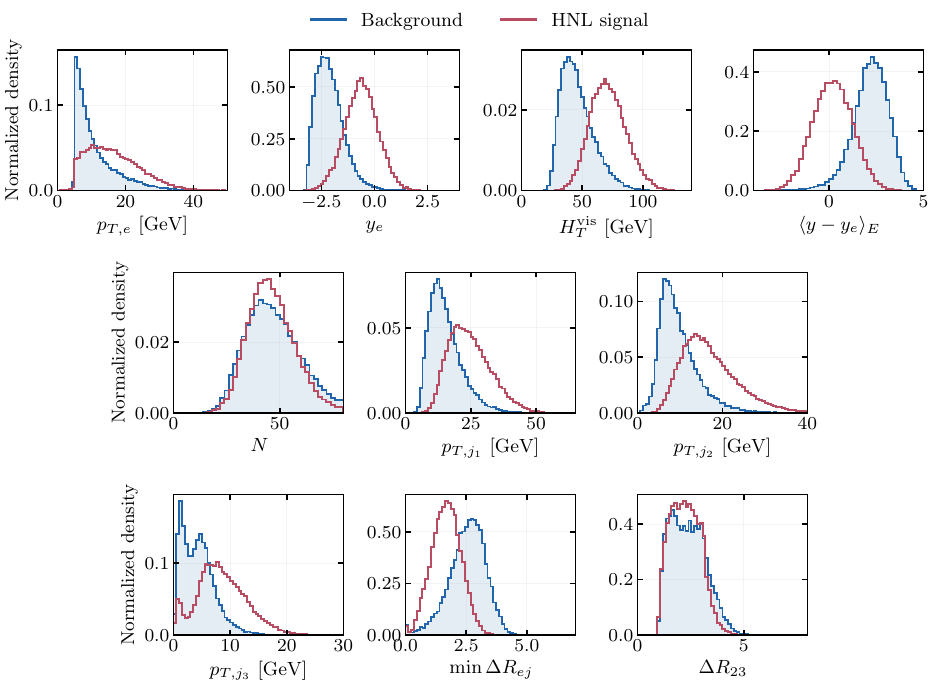}
 \caption{Normalized HNL signal and background distributions after laboratory acceptance and matching in 0.25~GeV bins of the SR mass $\mhnl$. Each class contains 32,683 events. The observables describe the electron, visible transverse activity, particle multiplicity, and electron-jet and jet-jet geometry.}
 \label{fig:hnl-physics}
\end{figure}

Figure~\ref{fig:hnl-physics} shows the corresponding event kinematics after mass matching. The electron rapidity, visible transverse activity, and energy-weighted mean rapidity displacement separate the classes strongly. The electron also tends to be closer to the candidate decay jets. Multiplicity and the separation of those jets are more similar. As in the $B-L$ example, the useful information comes primarily from resolved kinematics. For $B-L$ it reflects recoil against a radiated resonance, while for HNL it reflects the decay of a heavy particle.

\begin{figure}[tp]
 \centering
 \includegraphics[width=.98\linewidth]{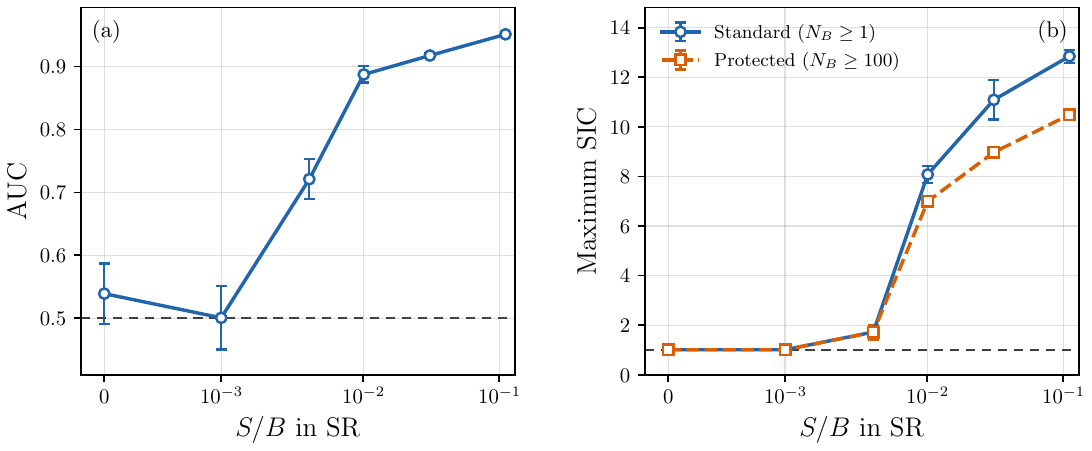}
 \caption{HNL classifier performance versus $S/B=f_S/(1-f_S)$. Left: signal-versus-background AUC. Right: maximum SIC, evaluated over thresholds with at least one surviving test-background event and over the restricted range with at least 100. Points are means over three independent five-fold trainings, and error bars show the standard error of the mean. The scan contains $f_S=0$, 0.001, 0.004, 0.01, 0.03, and 0.10. Each seed is evaluated on 40,000 signal and 100,000 background events. The horizontal reference lines indicate chance-level AUC and unit SIC. }
 \label{fig:hnl-fractions}
\end{figure}

\begin{figure}[tp]
 \centering
 \includegraphics[width=.72\linewidth]{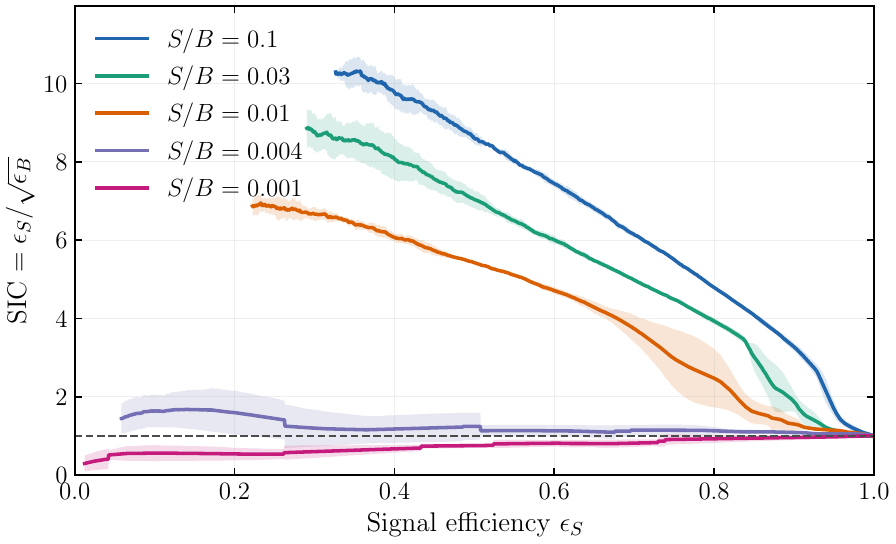}
 \caption{HNL significance-improvement curves for the nonzero signal fractions in figure~\ref{fig:hnl-fractions}, using PET2 with PID enabled. For each seed, the classifier scores from the five held-out folds are combined. Lines show the mean SIC at each signal efficiency over three seeds, and bands show one standard deviation. Curves are shown only where at least 100 test-background events survive in every seed. Legend values give $S/B=f_S/(1-f_S)$.}
 \label{fig:hnl-sic-extra}
\end{figure}

Figure~\ref{fig:hnl-fractions} shows how the full-event classifier becomes sensitive to the HNL signal as its fraction in the training sample increases. In the left panel, the background-only sample and the smallest signal admixture give no reliable separation, while larger admixtures allow the classifier to learn the differences between signal and background. The right panel shows the resulting signal enrichment. The requirement of at least 100 surviving test-background events has a smaller effect on the maximum SIC than in the $B-L$ benchmark. Much of the gain, therefore, remains when the background efficiency is estimated from a more populated sample. Figure~\ref{fig:hnl-sic-extra} further shows that, once the training becomes effective, useful signal enrichment is achieved over a broad range of signal efficiencies. These results demonstrate that the distinctive electron and jet kinematics of the HNL decay can be learned from mixed samples without identifying individual signal events during training.

\begin{figure}[!htbp]
 \centering
 \includegraphics[width=\searchFigureWidth,trim=0 20bp 0 0,clip]{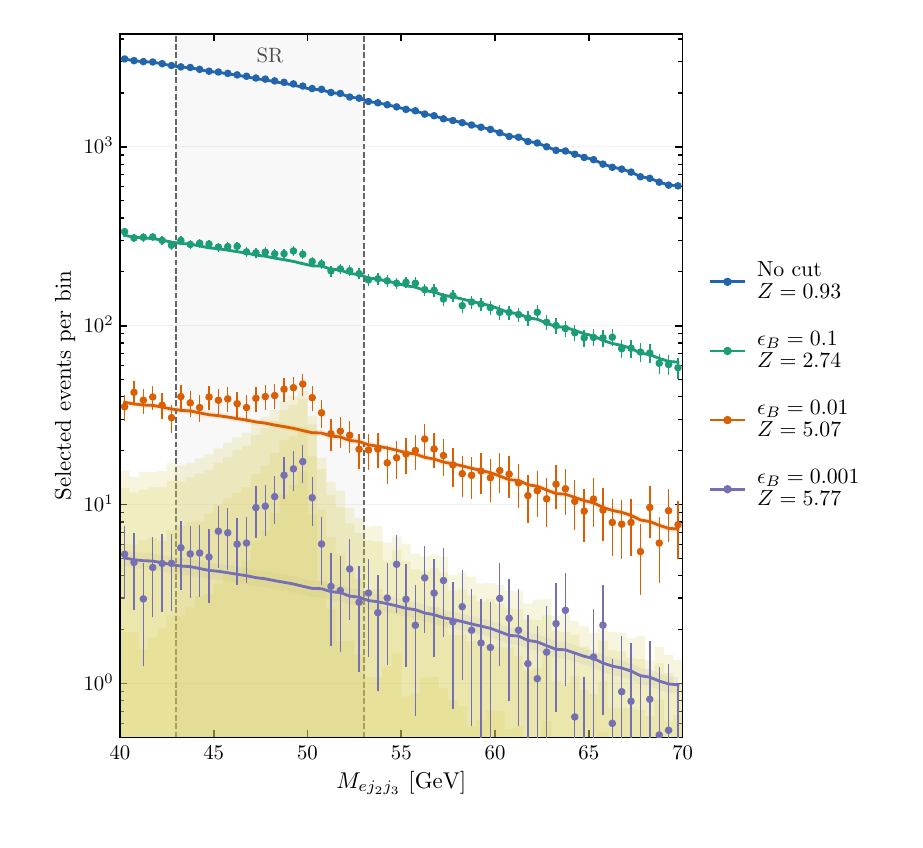}
 \caption{HNL mass spectra before and after classifier selection at $f_S=0.01$. Points show the mixed sample, continuous curves show the background prediction normalized in the sidebands, and yellow histograms show the signal contribution. Vertical lines and shading delimit the SR, $43<\mhnl<53$~GeV. All selections use the same overall event normalization and one five-fold classifier training. The quoted $Z$ values are expected local significances computed with the likelihood of section~\ref{sec:inference}, including finite reference statistics.}
 \label{fig:hnl-extra}
\end{figure}

Figure~\ref{fig:hnl-extra} shows how this discrimination improves the search for a resonance in the invariant mass of the electron and two candidate decay jets. The classifier suppresses the continuum background while retaining a substantial part of the resonant contribution, making the excess more prominent as the selection is tightened. For this benchmark, the tightest selection improves the expected local significance by approximately a factor of six at the same overall event normalization. The HNL result thus extends the demonstrated sensitivity of the method to a resonance whose reconstructed decay includes the identified electron, complementing the hadronic resonance studied in the $B-L$ benchmark.

\section{Conditional generation of the background reference}
\label{sec:generator}

\subsection{Full-event generation from sidebands}

The independent simulated background used so far provides an ideal reference that is unavailable in an experimental search. To apply the method to data, we instead need to learn a background reference from mass sidebands, following the conditional-generation procedure sketched in figure~\ref{fig:workflow}(b). As a proof of concept, we focus on the $B-L$ benchmark, but we expect that the same strategy can be extended to the HNL benchmark and other resonant searches.  Conditional density estimation and generation are widely used for resonant anomaly detection~\cite{Nachman:2020lpy,Hallin:2021wme,Raine:2022hht}, including generation of the full particle phase space~\cite{Buhmann:2023pmh}. Generative studies of EIC events \cite{Devlin:2023jzp,Araz:2024bom} and heavy-ion collisions~\cite{Go:2024hwi,Sadek:2026fkr} provide further motivation for extending these methods to complete events.

We construct the event in stages, following the sequential event-generation strategy of Ref.~\cite{Araz:2024bom}. At a requested resonance mass $m$, we first generate the scattered electron, then the overall momenta and particle content of the hadronic system, and finally the individual non-electron particles. The mass $m$ specifies the leading-dijet invariant mass after the analysis acceptance and jet reconstruction for the $B-L$ scenario. It is the variable along which the model must interpolate from the sidebands into the signal region.

We describe the electron by the vector ${\bf u}_{B-L}$ defined in equation~\eqref{eq:globals-bl}, which contains its transverse momentum, rapidity, and azimuth together with $x_B$ and $Q^2$. We sample the electron momentum and calculate the DIS variables from it using the relations in section~\ref{sec:observables}. Thus the six entries represent one electron and its associated DIS kinematics. We require $Q^2>0$, $0<x_B<1$, and $0<y_{\rm DIS}=Q^2/(s x_B)<1$, where $y_{\rm DIS}$ is the DIS inelasticity. Proposals that fail any of these conditions are redrawn at the same requested mass before generating the rest of the event.

The next stage generates an intermediate description of the visible non-electron particles. We divide this collection into four components: its three leading anti-$k_T$, $R=1$ jets in the center-of-mass frame and all remaining particles. For each component $c$, the model samples a total four-vector $P_c^\mu$ and counts $n_{c,a}$ in 17 particle-species and charge categories. We denote these conditions by
\begin{equation}
 S=\{P_c^\mu,n_{c,a}\},\qquad c=1,\ldots,4,\quad a=1,\ldots,17.
 \label{eq:genconditions}
\end{equation}
The multiplicity is $N_c=\sum_a n_{c,a}$. The four-vectors specify the energy and momentum of each component, while the counts specify its particle content. These components are defined before laboratory acceptance. The analysis jets are reconstructed later from the accepted particles and can therefore differ from the intermediate components.

Writing $\mathcal X$ for the non-electron particle collection, we express the event proposal as
\begin{equation}
 q_{\rm prop}(\mathcal X,{\bf u}_{B-L}\mid m)=\int dS\,
 q_{\rm event}({\bf u}_{B-L},S\mid m)\,
 q_{\rm particles}(\mathcal X\mid {\bf u}_{B-L},S,m).
 \label{eq:genfactor}
\end{equation}
The first factor generates the electron and component conditions, and the second generates the particles given those conditions. The integral includes sums over the discrete counts. In this way, the model generates both the overall event configuration and the particles that populate it. The particles in all four components are generated jointly, allowing the model to learn correlations between different jets and particles outside them. Reproducing these correlations is necessary because the search classifier uses the full event.

The particle generator uses the generative PET2 architecture and pretrained OmniLearned weights~\cite{Bhimji:2025isp}. We retain the pretrained attention layers, flow-time representation, and compatible generative layers. The layers that encode particle coordinates, species, local particle interactions, and event conditions are newly initialized for the EIC representation, as is the final coordinate-output layer. The adapted network has approximately 2.13 million parameters, including 1.45 million transferred parameters, which are fine-tuned together with the new layers. The auxiliary networks that generate the electron, component conditions, and photon pairing are trained on EIC sidebands without transferred weights.

To extend the single-jet generator to a full event, we replace its jet-level conditioning with the requested mass, the electron vector ${\bf u}_{B-L}$, and the momenta and multiplicities of the four components in $S$. Each particle also carries its species, charge, component identity, and the conditions of its component. One attention network processes all four components together. Following the OmniLearned generation procedure, it transforms noise into particle coordinates consisting of relative rapidity, relative azimuth, and a logarithmic momentum variable. Jet particles are represented relative to their jet axis, while particles outside the three jets use the electron direction as their reference. The particle interactions are evaluated in a common event frame, preserving the physical angular separations between particles in different components.

A separate model samples photon pairs treated as neutral-pion candidates, including pairs within and between components. Its training labels identify unambiguous pairs with invariant mass close to the neutral-pion mass. Additional interactions between paired photons allow their momenta to be generated jointly. The particle factor in equation~\eqref{eq:genfactor} includes a sum over the sampled pair assignments.

\subsection{Training and kinematic consistency}

All generator components are fitted to background sidebands in $20<\mjj<34$~GeV and $44<\mjj<60$~GeV. We also withhold $28<\mjj<32$~GeV for an interpolation check. No signal-region event is used to fit the generator. The electron/DIS network uses 5 million training events and 300,000 validation events. The count, four-vector, and main particle models each use one million training events and 100,000 validation events. Input normalization is determined from training events only.

During particle training, the electron and component conditions are taken from the same sideband event as the target particles. We use the OmniLearned training objective, giving each occupied component equal weight so that components with many particles do not dominate. The network learns from examples how the constituent momenta are related to the component totals. We impose no momentum-conservation or photon-pair-mass penalty during training. Instead, we enforce these constraints after generation, in a separate step that does not enter the training gradients.

We adjust the generated three-momenta so that the particles in each component sum to its sampled four-vector. Each particle retains its species mass, and its energy is recomputed from its adjusted momentum. Sampled photon pairs must also reproduce the neutral-pion invariant mass. The numerical solver applies successive weighted least-squares corrections that favor small relative momentum changes. Proposals that fail the component four-vector or pair-mass checks are rejected.

These constraints match the sampled component totals. They allow nonzero visible recoil because invisible particles are excluded, and they impose no momentum balance within the detector acceptance. After the adjustment, we apply the common laboratory acceptance and reconstruct the analysis jets in the center-of-mass frame. We retain events with $|\mjj^{\rm reco}-m|<0.25$~GeV so that the reference corresponds to the requested mass. 

\subsection{Sideband-derived probability correction and validation}

The accepted generated events can still differ from the background. We use a separate classifier to identify such differences in the sidebands and assign weights to the generated events. Configurations that are underrepresented receive larger weights, while overrepresented configurations receive smaller weights. This correction changes their contribution to the reference sample while leaving their particle momenta unchanged.

The correction classifier distinguishes background sideband events from generated events at the same reconstructed mass $m_{\rm reco}$. Here $x$ denotes the event inputs used by the classifier, and $b(x\mid m_{\rm reco})$ and $q(x\mid m_{\rm reco})$ are their conditional densities for background and accepted generated events. We use the reconstructed mass after selection for this comparison, rather than the requested mass of the initial proposal. Let $D_{\rm corr}(x,m_{\rm reco})$ be the probability assigned to the background class. For equally populated training classes, the classifier odds estimate the density ratio,
\begin{equation}
 w(x,m_{\rm reco})=\frac{D_{\rm corr}(x,m_{\rm reco})}{1-D_{\rm corr}(x,m_{\rm reco})}
 \simeq\frac{b(x\mid m_{\rm reco})}{q(x\mid m_{\rm reco})}.
 \label{eq:reweight}
\end{equation}
The weights define the corrected reference,
\begin{equation}
 \widetilde q(x\mid m_{\rm reco})=
 \frac{w(x,m_{\rm reco})q(x\mid m_{\rm reco})}{\int dx'\,w(x',m_{\rm reco})q(x'\mid m_{\rm reco})}.
\end{equation}
The denominator normalizes the distribution at each reconstructed mass.

Classifier-based density-ratio estimation and event reweighting are well established. See, for example, Refs.~\cite{Cranmer:2015likelihood, Rogozhnikov:2016reweight,Andreassen:2019reweight}. We use two correction classifiers, both trained to distinguish background sideband events from generated events. Their combination gives a lower classification loss on held-out sideband events than either classifier alone. We calibrate their outputs and fit a linear combination of the logarithms of their output odds using a separate sideband sample. Exponentiating this combination gives the event weights, which we normalize within reconstructed-mass bins. These weights enter the training and validation losses of the PET2 search classifier. The correction compares background with generated events, while the search compares the $S+B$ mixture with the corrected reference. For this proof of concept, we simplify both classification tasks by omitting explicit PID and charge inputs. The generator still samples particle species to construct physical final states. The correction is learned from background sidebands without using labeled signal events. In an experimental search, it would need to interpolate from background-dominated data sidebands into the excluded mass window. We assess the final reference after both the kinematic adjustment and the probability correction.

\begin{figure}[tp]
 \centering
 \includegraphics[width=.98\linewidth]{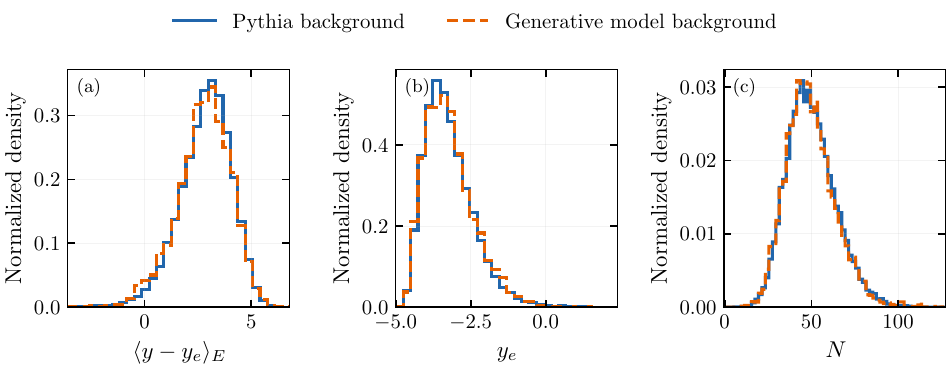}
\caption{Quality checks of the corrected generative-model background against an independent \PYTHIA\ sample in the signal region. The panels show (a) the energy-weighted rapidity displacement $\langle y-y_e\rangle_E$, (b) electron rapidity, and (c) accepted non-electron multiplicity, as defined in section~\ref{sec:observables}. The reconstructed-mass distributions are matched, and each panel is normalized to unit area.}
 \label{fig:generator-extra}
\end{figure}

Figure~\ref{fig:generator-extra} compares representative observables of the corrected reference with an independent \PYTHIA\ sample in the signal region, after the same acceptance and jet reconstruction, and with matched reconstructed-mass distributions. The observables probe the electron kinematics, the location of the hadronic energy relative to the electron, and the accepted particle multiplicity. These distributions test aspects of the event beyond the agreement in reconstructed mass imposed by the selection. The model reproduces their main features, with small residual differences. In background control regions, a separately trained classifier provides a more sensitive check using the combined event inputs. We evaluate it on held-out events with matched mass distributions and compare it with a control using two independent \PYTHIA\ samples. The model captures the main features of the displayed distributions, while this multivariate check identifies residual differences in their correlations. The search comparison below demonstrates that the resulting reference supports substantial signal enrichment and measures the remaining difference in sensitivity relative to the ideal reference.

\subsection{Impact on significance improvement}

We now test how replacing the ideal reference by the learned background affects signal enrichment. At each signal fraction, we train a pair of PET2 classifiers on the same $S+B$ events, using either a \PYTHIA\ background reference or the corrected generative-model background. The paired fits start from the same pretrained parameters and use matched reconstructed-mass distributions and the same evaluation sample. Both retain the electron/DIS inputs and omit explicit species and charge inputs. The generator and correction remain fixed throughout.

The scan uses the same signal fractions as the ideal-background study in section~\ref{sec:results}. For this proof-of-concept comparison, we omit explicit PID and charge inputs and use 15,000 training and 3,000 validation events per weak class. We apply these settings to both background references, so the ideal-background curve provides a comparison with matched inputs and sample sizes rather than reproducing the performance in section~\ref{sec:results}. We repeat each comparison for three choices of training sample and initialization, and evaluate every fit on the same held-out sample of 20,000 signal and 20,000 \PYTHIA\ background events.

\begin{figure}[tp]
 \centering
 \includegraphics[width=.68\linewidth]{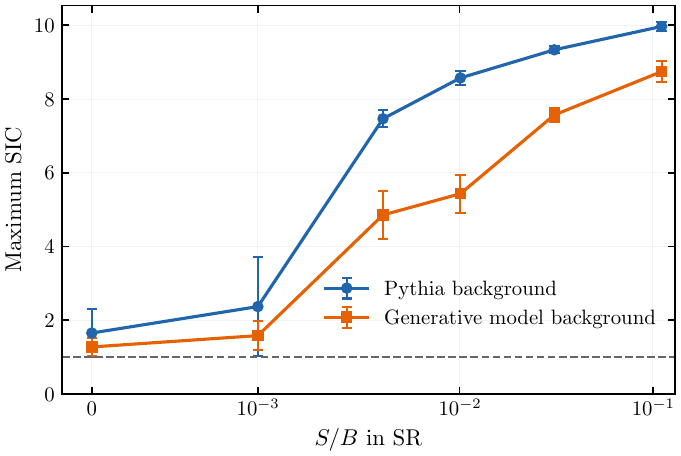}
 \caption{Maximum SIC for the $B-L$ benchmark using a \PYTHIA\ background reference or a generative-model background with sideband-derived correction. The signal fractions are plotted as $S/B=f_S/(1-f_S)$ in the SR. The paired fits use the same PID-blind PET2 classifier, mixed-event sample, and held-out evaluation. Each fit uses 15,000 training and 3,000 validation events per weak class. Points and error bars show the mean and standard error over three classifier trainings, with the generator and correction fixed. The maximum requires at least 100 of the 20,000 test-background events to pass. The dashed line marks unit SIC.}
 \label{fig:generator-sic}
\end{figure}

Figure~\ref{fig:generator-sic} shows the maximum SIC, calculated separately for each fit with at least 100 surviving test-background events and then averaged. The error bars describe the variation across the three trainings of the classifier. They do not include the uncertainty of fitting the generator or interpolating across the mass window.

The learned reference preserves substantial signal enrichment once the signal admixture is large enough for the training to be effective. Its performance remains below that obtained with the ideal reference, with a smaller relative difference at the larger signal fractions. Residual differences between the two backgrounds compete with the signal during training and reduce the resulting enrichment. The generated reference still supports a useful search in this benchmark, providing a proof of concept for learning the background from sidebands.

At the smallest nonzero signal fraction, the results vary strongly between trainings. The background-only training point is also evaluated on the held-out signal and background samples, so its measured maximum SIC can exceed one. It tests signal separation without an injected training signal and does not represent a background-only discovery significance.

The generated reference supports useful signal enrichment in this study. However, with larger data samples, the classifier can become sensitive to smaller differences between the generated and true backgrounds. A separate comparison with larger classifier samples gives less enrichment, although changes in training and evaluation settings prevent attributing this solely to sample size.

To illustrate the role of background accuracy, write the corrected background density at fixed reconstructed mass as $\widetilde q(x)=b(x)[1+\delta(x)]$, where $\delta(x)$ describes the relative modeling error. Using the densities defined in section~\ref{sec:learning}, the optimal weak classifier ranks events by
\begin{equation}
 \frac{p_{\rm mix}(x)}{\widetilde q(x)}=
 \frac{1+f_S[s(x)/b(x)-1]}{1+\delta(x)}.
 \label{eq:gencompetition}
\end{equation}
The numerator contains the signal information, whose contribution becomes smaller as the signal fraction decreases. The denominator describes the background differences that the classifier can also learn. Improving the background model therefore offers a route to exploiting larger data samples and smaller signal fractions. Generating more events reduces sampling fluctuations, but improving the model itself is necessary to reduce $\delta(x)$. The appropriate background accuracy should consequently be assessed for the sample size and selections of the intended search.

The comparison measures the sensitivity loss for one fitted generator. We leave a more quantitative study of uncertainties from generator training and sideband interpolation, including the dependence on generator and classifier sample sizes, to future work. An experimental application would also require adapting the model to reconstructed particles within detector acceptance.
\section{Fully unsupervised graph autoencoder}
\label{sec:ae}

\subsection{Event reconstruction and anomaly score}

An autoencoder provides a complementary strategy when the signal fraction is too small to learn from weak labels. It compresses an event into a lower-dimensional representation and reconstructs the input from that latent representation. This implicitly biases the autoencoder to focus on properly reconstructing the most typical events in the dataset, potentially leading to a high reconstruction loss for the rare BSM events in the dataset. We train it on a background-dominated sample and use the reconstruction error as an anomaly score. The training does not require separation into signal-enriched and reference classes. Its sensitivity depends on whether the signal is harder to reconstruct than the background~\cite{Finke:2021sdf,Fraser:2021lxm}.

Here, we use a sparse relational graph autoencoder for complete $B-L$ events, extending the jet-graph approach of Ref.~\cite{Araz:2026graph}. Graph and set models naturally accommodate the variable number of particles~\cite{Atkinson:2021nlt,Ostdiek:2021bem,Tsan:2021brw,Cheng:2020dal}. Each accepted particle, including the identified electron, is a node. Edges encode selected pairwise relations through which the network exchanges information before reconstructing both the nodes and their relations.

We represent each event by a graph $G=(V,E)$, where $V$ is the set of $N=|V|$ particle nodes and $E$ contains both directions of each selected connection between particles. For node $i$, the input vector is $v_i=(p_{Ti},\eta_i,\phi_i,\mathrm{PID}_i)$, where $\eta_i$ is the COM-frame pseudorapidity and the signed PDG identifier is stored numerically. For an edge connecting $i$ and $j$, we use the variables $a_{ij}=(\ln\Delta_{ij},\ln k_{T,ij},\ln z_{ij})$, where
\begin{align}
 \Delta_{ij}&=[(\Delta\eta_{ij})^2+(\Delta\phi_{ij})^2]^{1/2},\\
 k_{T,ij}&=\min(p_{Ti},p_{Tj})\,\Delta_{ij},\qquad
 z_{ij}=\frac{\min(p_{Ti},p_{Tj})}{p_{Ti}+p_{Tj}}.
\end{align}
Here $\Delta_{ij}$ is the geometric distance in the $\eta-\phi$ plane using the shortest periodic azimuthal separation. These edge variables describe angular separation, relative transverse scale, and momentum sharing. For the full-event EIC benchmark, the highest expected resonance significance among the tested configurations was obtained when retaining absolute angular coordinates, whereas relative-position inputs performed better in the LHC jet study~\cite{Araz:2026graph}.

Following Ref.~\cite{Araz:2026graph}, we use sparse unique-$5$ graphs. Particles are sorted by decreasing $p_T$, and the five hardest form a fully connected subgraph. Each subsequent particle connects to its five nearest harder predecessors in the periodic $(\eta,\phi)$ plane. Information passes in both directions along these connections. The construction imposes no partition into jets, so it includes relations across the event. In a direct comparison, unique-$5$ graphs give a higher AUC and maximum SIC than fully connected graphs. Increasing the background training sample does not improve either metric for the two graph constructions in the configurations tested. We retain unique-$5$ graphs for all the results presented below.

Denoting reconstructed quantities by hats, we define the reconstruction score for an individual event from the mean squared errors in its node and edge features,
\begin{align}
 \mathcal L_{\rm AE}&=\mathcal L_{\rm node}+\lambda\mathcal L_{\rm edge},
 \qquad\lambda=0.6,\label{eq:ae}\\
 \mathcal L_{\rm node}&=\frac{1}{4N}\sum_{i=1}^{N}
 \sum_{\alpha=1}^{4}(\widehat v_{i\alpha}-v_{i\alpha})^2,\qquad
 \mathcal L_{\rm edge}=\frac{1}{3|E|}\sum_{(i,j)\in E}
 \sum_{\beta=1}^{3}(\widehat a_{ij,\beta}-a_{ij,\beta})^2.
 \label{eq:ae-components}
\end{align}
The first term measures how well the model reconstructs each particle, and the second measures how well it reconstructs the selected particle pairs. Their weighted sum gives the event score. During training, the same squared residuals are averaged over all nodes and all edges in the minibatch, respectively. This differs from assigning equal weight to each event's reconstruction score. This implementation retains the absolute angular coordinates of each particle as well as the pairwise features. Particle identity enters the reconstruction loss through its numerical PDG value, whereas the classifier learns a categorical representation.

\subsection{Dependence on the signal fraction and complexity bias}
\label{sec:ae-interpretation}

We first examine how the reconstruction score depends on the signal admixture in the training sample. We train on mixed events in $20<\mjj<60$~GeV, with 200,000 background events for training and 25,000 for validation. We add signal events to obtain each of the signal fractions used in section~\ref{sec:results}, defined within the SR. The signal contribution outside the SR follows its mass distribution. We repeat each training for five random seeds, keeping the graph construction, loss, and training duration of 100 epochs fixed.

\begin{figure}[tp]
 \centering
 \includegraphics[width=.98\linewidth]{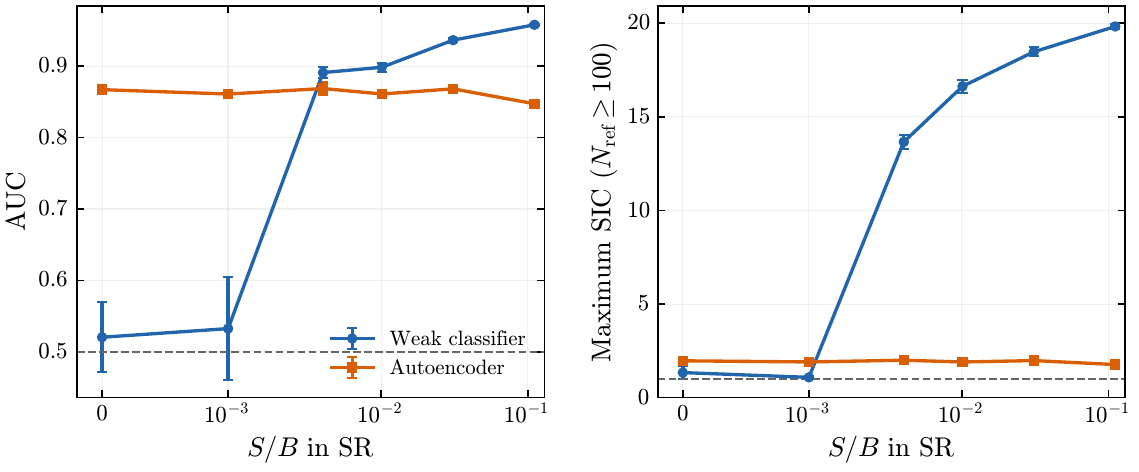}
 \caption{$B-L$ signal-versus-background AUC (left) and maximum SIC (right) for the graph autoencoder and weakly supervised classifier, versus $S/B=f_S/(1-f_S)$ in the SR. The classifier results are those of figure~\ref{fig:bl-fractions}. Both methods are evaluated on the same 100,000 signal and background events. Each event is scored by a model that did not use it for training or validation. The maximum SIC requires at least 100 surviving test-background events. Points and error bars show means and SEM over five autoencoder trainings and three five-fold classifier trainings, respectively. The dashed lines indicate chance-level AUC and unit SIC.}
 \label{fig:ae-fractions}
\end{figure}

Figure~\ref{fig:ae-fractions} compares the resulting AUC and maximum SIC with the weakly supervised classifier. Both methods are evaluated on the same held-out signal and background events in the SR. The autoencoder retains useful separation even without signal contamination in its training sample, and its performance changes little over the smaller signal fractions. Weak supervision instead needs enough signal events to learn the difference between the mixed sample and the reference. Once this difference can be resolved, it gives a larger significance improvement, which grows with the signal fraction. The comparison illustrates the complementary regimes of the two approaches~\cite{Collins:2021nxn}. Reconstruction can provide a signal enhancement when a small admixture is insufficient for weak supervision, while classification can exploit differences in event frequencies once the admixture is large enough.

%\subsection{Reconstruction bias and event kinematics}

The comparison with weak supervision shows that the full event contains discriminating information that is only partly captured by the reconstruction score. For the $B-L$ benchmark, signal and background have similar particle content, but different distributions of electron and jet momenta and their correlations, as illustrated in figure~\ref{fig:bl-physics}. A configuration can occur more frequently in signal without being difficult for a background-trained network to reconstruct. Weak supervision can exploit these differences in frequency, whereas the autoencoder selects events according to the size of their reconstruction errors.

\begin{figure}[tp]
 \centering
 \includegraphics[width=.68\linewidth]{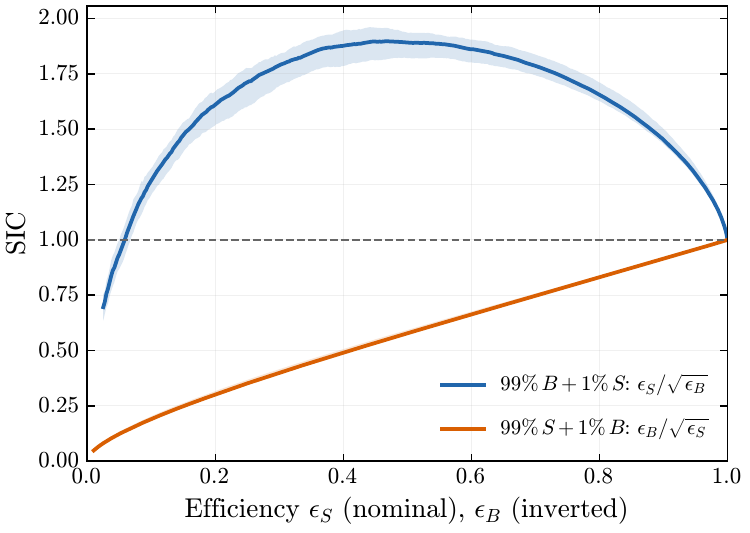}
 \caption{Training-sample inversion for the $B-L$ graph autoencoder. The nominal sample contains one percent signal in background, while the inverted sample contains one percent background in signal, with both fractions defined in the SR. Selecting larger reconstruction errors gives $\epsilon_S/\sqrt{\epsilon_B}$ versus $\epsilon_S$ in the nominal case and $\epsilon_B/\sqrt{\epsilon_S}$ versus $\epsilon_B$ in the inverted case. Curves show means over four trainings and bands span their minimum and maximum. Both use the same held-out signal and background events as figure~\ref{fig:ae-fractions}. At least 100 reference events must survive in every seed, where the reference is background for nominal training and signal for inverted training. The dashed line marks unit SIC.}
 \label{fig:ae-inversion}
\end{figure}

Figure~\ref{fig:ae-inversion} tests whether large reconstruction errors identify the rare component when the training populations are exchanged. In the nominal case, we train on predominantly background sample with one percent signal in the SR. In the inverted case, we train on predominantly signal with one percent background in the SR. The dominant class supplies the same numbers of training and validation events in both cases. We keep the architecture and training procedure fixed and compare four random seeds in each direction, always selecting larger reconstruction errors.

The nominal selection improves sensitivity to signal, but the inverted selection gives no significance improvement for the injected background. Thus, reconstruction error does not simply follow rarity in the training sample. One possible interpretation is that the range of configurations a model reconstructs accurately extends beyond the dominant training population. A signal-trained model may also reconstruct softer or simpler background configurations well, while a background-trained model need not reconstruct signal configurations equally well. This asymmetry can therefore reflect the kinematic range learned by the model as well as differences in structural complexity. It is consistent with the reconstruction bias observed in LHC studies. Since similar behavior occurs in both settings, the inversion test alone does not explain the smaller significance improvement obtained here.

This preference for assigning larger reconstruction errors to signal does not by itself guarantee a large gain in search sensitivity. The gain depends on how much background remains in the high-error tail relative to the signal retained by the selection. Here, the autoencoder separates typical signal and background events, but this separation does not persist as strongly in the high-error tail. Its significance improvement peaks at intermediate selections and falls below unity at the tightest selections shown. Beyond this maximum, tightening the cut removes signal too rapidly for the additional background rejection to compensate. A high AUC alone does not capture this limitation. These results suggest that the present reconstruction objective does not fully exploit the discriminating information in the event kinematics. They motivate developing representations and objectives better suited to correlations across the full electron-hadron final state~\cite{Buss:2022lxw,Dillon:2021nxw,Dillon:2022mkq}.

\subsection{Autoencoder selection in the resonance search}

\begin{figure}[tp]
 \centering
 \includegraphics[width=\searchFigureWidth,trim=0 20bp 0 0,clip]{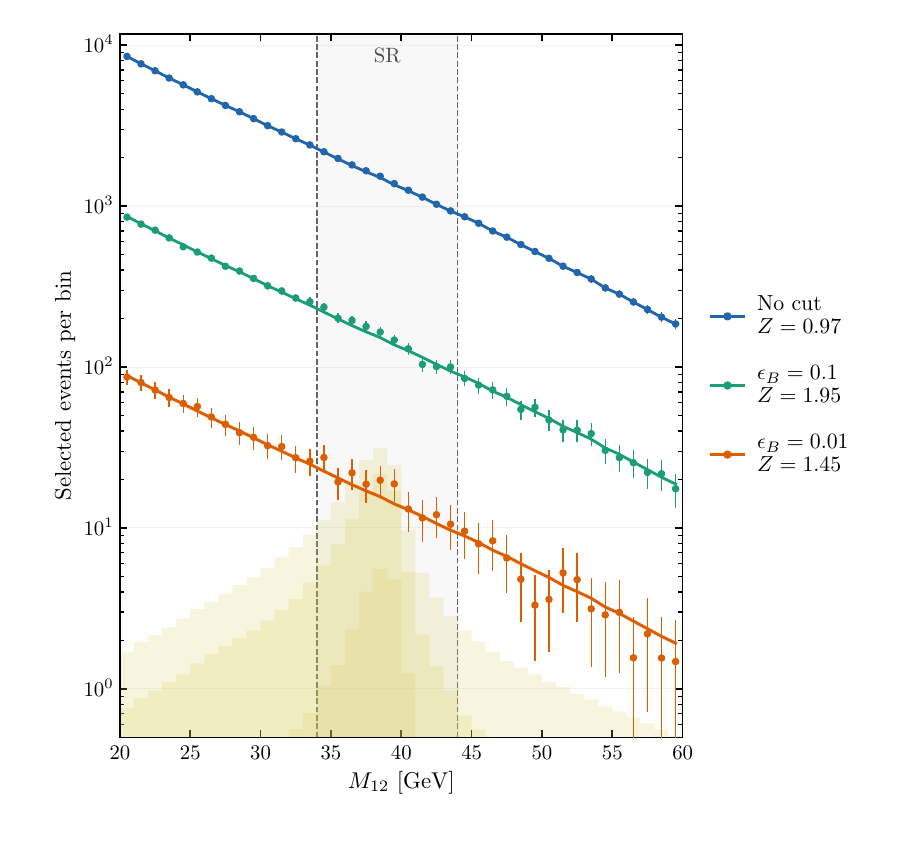}
 \caption{$B-L$ mass spectra after graph-autoencoder selection at $f_S=0.01$ with $|\eta_{\rm lab}|<4$ acceptance. The five-model ensemble is trained on a mixed sample with the same signal fraction, using only reconstruction loss. Points show the selected mixed sample, curves the reference prediction, and histograms the signal contribution. All working points use a common exposure. The quoted $Z$ values are expected local significances evaluated with the finite-reference likelihood.}
 \label{fig:ae-bump-extra}
\end{figure}

We finally test whether the reconstruction score can enhance the sensitivity of a resonance search. Figure~\ref{fig:ae-bump-extra} shows the mass spectra before and after selection at $f_S=0.01$, using the same overall event normalization as the classifier comparison. For this search, we combine five models based on unique-$5$ graphs trained on a background-dominated sample with the same signal contamination. A moderate selection approximately doubles the expected local significance, demonstrating that the autoencoder can enhance an excess that is difficult to identify from the mass spectrum alone.

Together, the fraction scan and resonance search show that fully unsupervised reconstruction provides useful sensitivity even when the training sample contains very little signal. For the present benchmark, weak supervision gives a larger gain once enough signal is available to learn the differences in event frequencies. The informative electron and resolved hadronic final state at the EIC offer a setting for developing anomaly scores that exploit these correlations while retaining sensitivity at small signal fractions.
\FloatBarrier

\FloatBarrier
\section{Conclusions and outlook~\label{sec:conclusions}}

The observed matter-antimatter asymmetry and the nature of dark matter and dark energy, together with other indications that the Standard Model is incomplete, motivate searches for new interactions, but leave their experimental signatures largely undetermined. Anomaly detection complements dedicated searches for specific models by seeking deviations from the expected background without requiring a model for the signal distribution. The EIC offers opportunities to develop this approach through the information carried jointly by the scattered electron and the hadronic final state. New interactions can change their momentum distributions and correlations even when the observed particles resemble those produced in ordinary deep-inelastic scattering.

We have studied full-event anomaly detection using a $B-L$ gauge boson and a heavy neutral lepton as representative resonances. In the $B-L$ case, the scattered electron recoils against a hadronically decaying resonance. In the production of HNL, the electron itself is a decay product of the BSM particle. Weakly supervised classification exploits the resulting differences in event kinematics to enhance both resonance searches, without identifying individual signal events during training. For the $B-L$ benchmark with an ideal background reference and $S/B = 0.01$, the expected local significance increases from approximately one to ten. This order-of-magnitude improvement demonstrates that event information can reveal a resonance that would otherwise remain insignificant. The comparison with summary statistics shows that electron kinematics provide substantial sensitivity, while the full-event representation further improves on the chosen summaries.

Applying this strategy to experimental data requires learning the background from control region measurements. For the $B-L$ benchmark, we have constructed a conditional model that generates complete events from the sideband regions, including correlations between the electron, jets, and remaining particles. The generated background supports substantial signal enrichment across the scan of signal fractions. This provides a proof of concept for extending sideband-based resonance searches to the full event. The comparison with an ideal reference measures the loss in sensitivity from background modeling and identifies improved reference accuracy as a route toward sensitivity to smaller signal admixtures.

We also studied a graph autoencoder, which provides a complementary approach without requiring a separate reference class during training. This approach is particularly relevant when the signal fraction is too small for weak supervision to learn a useful separation. Its reconstruction score enhances the resonance, although with a more modest gain than weak classification in the comparison considered here. The results highlight the distinction between an unusual event configuration and an unusual frequency of otherwise familiar configurations. A new interaction can change the latter without making individual events harder to reconstruct. Weak classification can exploit these differences in frequency, while reconstruction error does not need to be sensitive to them. These findings motivate the development of learning objectives that better capture the changes in event kinematics and correlations through which new physics could appear at the EIC.

An experimental application will require inclusion of the detector response, systematic uncertainties, and significance calibration for a scan over unknown resonance masses. For the learned background, an important next step is to quantify uncertainties from generator training and sideband interpolation, together with their dependence on the data available to the generator and search classifier. These studies will establish how the demonstrated gains extend to a realistic experimental search.

The use of jet-pretrained networks for both classification and particle generation also motivates dedicated EIC pretraining. Together with event representations and learning objectives developed for electron-hadron collisions, this could improve the use of information throughout the final state. Beam polarization and nuclear targets offer additional ways to probe new interactions and investigate a possible excess. Developing these methods alongside the EIC's precision QCD program can broaden its searches for physics beyond the Standard Model and make fuller use of the information recorded in each collision.

\section*{Code availability}

The event samples are publicly available on Zenodo at \url{https://doi.org/10.5281/zenodo.22904017}. The code and trained models can be obtained from the authors upon request.

\section*{Acknowledgments}

We thank Yeonju Go, Tommaso Rainaldi, and Keping Xie for helpful discussions. We acknowledge seed grant funding from Stony Brook University and Brookhaven National Laboratory. FR is supported by the DOE, Office of Science, Office of Nuclear Physics, Early Career Program under contract No. DE-SC0025881. This work was supported by the U.S. Department of Energy, Office of Science, Office of Nuclear Physics, InQubator for Quantum Simulation (IQuS)\footnote{\url{https://iqus.uw.edu}} under Award Number DOE (NP) Award DE-SC0020970 via the program on Quantum Horizons: QIS Research and Innovation for Nuclear Science\footnote{\url{https://science.osti.gov/np/Research/Quantum-Information-Science}} (SG). Support is also acknowledged from the U.S. Department of Energy, Office of Science, National Quantum Information Science Research Centers, Quantum Systems Accelerator (Award No. DE-SCL0000121). This work was also supported by the U.S. Department of Energy, Office of Science, Office of Nuclear Physics, Grant No. DE-FG02-97ER-41014 (UW Nuclear Theory). SG was supported in part by a Feodor Lynen Research fellowship of the Alexander von Humboldt foundation. SG thanks
the EIC theory institute at Brookhaven National Lab for its support and hospitality. RS and HL are supported by the United States Department of Energy under Grant Contract DE-SC0012704. We acknowledge the use of {\tt Codex} and {\tt Claude}. This research used resources of the National Energy Research Scientific Computing Center (NERSC), a DOE Office of Science User Facility using NERSC award NP-ERCAP0037787. The authors thank Stony Brook Research Computing and Cyberinfrastructure and the Institute for Advanced Computational Science at Stony Brook University for access to the SeaWulf computing system, made possible by National Science Foundation grants \#1531492 and Major Research Instrumentation award \#2215987, with matching funds from Empire State Development's Division of Science, Technology and Innovation (NYSTAR), contract C210148. This work was enabled, in part, by the use of advanced computational, storage and networking infrastructure provided by the Hyak supercomputer system at the University of Washington. This work was also supported, in part, through the Department of Physics\footnote{\url{https://phys.washington.edu}} and the College of Arts and Sciences\footnote{\url{https://www.artsci.washington.edu}} at the University of Washington.  

\bibliographystyle{JHEP}
\bibliography{references}
\end{document}